\documentclass[twocolumn]{aastex631}

\usepackage{xcolor}

\shorttitle{Impactful Encounters}
\shortauthors{Rose et al.}

\graphicspath{{./}{figures/}}

\begin{document}

\title{Stellar Collisions in the Galactic Center: \\ Massive Stars, Collision Remnants, and Missing Red Giants}

\correspondingauthor{Sanaea C. Rose}
\email{srose@astro.ucla.edu}

\author{Sanaea C. Rose}
\affiliation{Department of Physics and Astronomy, University of California, Los Angeles, CA 90095, USA}
\affiliation{Mani L. Bhaumik Institute for Theoretical Physics,
University of California, Los Angeles,
CA 90095, USA}

\author{Smadar Naoz}
\affiliation{Department of Physics and Astronomy, University of California, Los Angeles, CA 90095, USA}
\affiliation{Mani L. Bhaumik Institute for Theoretical Physics,
University of California, Los Angeles,
CA 90095, USA}

\author{Re'em Sari}
\affiliation{Racah Institute for Physics, The Hebrew University, Jerusalem 91904, Israel}

\author{Itai Linial}
\affiliation{Racah Institute for Physics, The Hebrew University, Jerusalem 91904, Israel}
\affiliation{Institute for Advanced Study, Einstein Drive, Princeton, NJ 08540, USA}

\begin{abstract}

Like most galaxies, the Milky Way harbors a supermassive black hole (SMBH) at its center, surrounded by a nuclear star cluster. In this dense star cluster, direct collisions can occur between stars before they evolve off the main-sequence. Using a statistical approach, we characterize the outcomes of these stellar collisions within the inner parsec of the Galactic Center (GC). Close to the SMBH, where the velocity dispersion is larger than the escape speed from a Sun-like star, collisions lead to mass loss. We find that the stellar population within $0.01$~pc is halved within about a Gyr because of destructive collisions. Additionally, we predict a diffuse population of peculiar low-mass stars in the GC. These stars have been divested of their outer layers in the inner $0.01$~pc before migrating to larger distances from the SMBH. Between $0.01$ and $0.1$~pc from the SMBH, collisions can result in mergers. Our results suggest that repeated collisions between lower mass stars can produce massive ($\gtrsim 10$~M$_\odot$) stars, and there may be $\sim 100$ of them residing in this region. We provide predictions on the number of G objects, dust and gas enshrouded stellar objects, that may result from main-sequence stellar collisions. Lastly, we comment on uncertainties in our model and possible connections between stellar collisions and the missing red giants in the GC.

\end{abstract}



\section{Introduction} \label{sec:intro}

Like most galaxies, the Mikly Way harbors a supermassive black hole (SMBH) at its center \citep[e.g.,][]{Genzel+03,Kormendy04,FerrareseFord05,Ghez+05,KormendyHo13}. A dense region of stars and stellar remnants, known as the nuclear star cluster, surrounds the SMBH \citep[e.g.,][]{schodel+03,Ghez+05,Ghez+08,Gillessen+09,Gillessen+17}.
The proximity of the Milky Way's Galactic Center (GC) presents a unique opportunity to observe the consequences of this dense, dynamic environment.
One such consequence
is that direct collisions can occur between stars \citep[e.g.,][]{Dale+09,DaleDavies,Mastrobuono-Battisti+21,Rose+20,Rose+22}. As we show below, any $1$~M$_\odot$ star within about $0.1$~pc of the SMBH experiences at least one collision over its lifetime. At $0.1$~pc from the SMBH, the collision timescale is comparable to the main-sequence lifetime of a $1$~M$_\odot$ star. Collisions are therefore essential to understanding the properties and demographics of the nuclear star cluster. 

A number of studies have explored the effects of this dynamical process on the nuclear star cluster.
For example, stellar collisions can modify the stellar density profile near the SMBH, where they deplete the stellar population \citep[e.g.,][]{DuncanShapiro83,Murphy+91,David+87a,David+87b,Rauch99,FreitagBenz02}.
Collisions have also been invoked to explain the dearth of $1$-$3$~M$_\odot$ red giants (RGs) in the GC, a persistent observational puzzle \citep[e.g.,][]{Genzel+96,BaileyDavies99,Buchholz+09,Do+09,Gallego-Cano+18,Habibi+19RG}. \citet{Genzel+96} first suggested that RG collisions with main-sequence stars may destroy the RGs. However, theoretical studies diverge on whether such a collision can divest a RG of its envelope, with more recent work indicating insufficient mass loss post-collision \citep[e.g.,][see the latter for discussion]{Alexander99,BaileyDavies99,Dale+09}. Other mechanisms, such as black hole-RG collisions, RG encounters with stellar binaries, and a nuclear jet from the SMBH during its active phase, can only partly explain the apparent underabundance of RGs within $0.1$~pc of the SMBH \citep[e.g.,][]{Davies+98,Dale+09,Zajacek+20}. The missing RGs therefore remains an open question, one which main-sequence stellar collisions may yet address \citep[see][and our discussion in Section~\ref{sec:RGs}]{Mastrobuono-Battisti+21}.

Collisions and mergers in the nuclear star cluster can also yield a wide range of observables \citep[e.g.,][]{AmaroSeoane23}. Stellar collisions may be associated with several electromagnetic signatures, including AGN variability \citep[e.g.,][]{Murphy+91,Torricelli-C+00,FreitagBenz02}, the presence of blue stragglers \citep[e.g.,][]{Sills+97,Sills+01,Lombardi+02}, and supernova-like explosions \citep{DaleDavies,Balberg+13}. Furthermore, stellar mergers represent a possible explanation for the origin of G objects, stellar objects that are enshrouded by gas and dust observed in the nuclear star cluster \citep[albeit secular binary mergers rather than impulsive collisions:][]{Antonini+10,Witzel+14,Witzel+17,Stephan+16,Stephan+19,Ciurlo+20}. Lastly, collisions between compact objects can release gravitational wave emission, potentially detectable by the LIGO-Virgo-KARAGA collaboration \cite[e.g.,][]{OLeary+09,Hoang+18,Hoang+20,ArcaSedda20}.

Also with implications for gravitational wave sources, repeated collisions between stars and black holes (BHs) can cause the latter to grow in mass, producing more massive BHs than predicted by stellar evolution models (see \citet{Rose+22}; for BH mass predictions, see also \citet{Heger+03,Woosley+17,SperaMapelli17,LimongiChieffi,Belczynski20,Renzo+20}). However, for the BHs to grow significantly, there must be a sufficiently large population of stars to sustain the collisions. Star-star collisions may deplete this supply, especially in the region closest to the SMBH, $\lesssim 0.01$~pc, where BH-star collisions can act most efficiently to produce larger BHs \citep{Rose+22}. Therefore, understanding the impact of collisions on the stellar population has important implications for the compact object-star interactions that give rise to massive BHs and interesting electromagnetic signatures (for the latter, see \citet{Kremer+22}). 

In this study, we consider the effect of stellar collisions on the masses and survival of stars in the GC, with implications for the topics described above. We show that collisions between low-mass stars can give rise to a population of young-seeming, massive stars in the vicinity of the SMBH as well as peculiar low-mass stars that have been divested of their outer layers. Additionally, our results suggest that stellar collisions may explain observations of dust-enshrouded gaseous stellar objects and a missing red giant cusp in the Galactic Center. This paper is organized as follows: In Section~\ref{sec:methodology}, we describe our methodology. Many studies rely on smooth particle hydrodynamics (SPH) to determine the outcome of a collision, a complex problem that depends on assumptions about the stellar structure, impact velocity, and mass ratio of the stars \citep[e.g.,][]{BenzHills87,Lai+93,Rauch99,FregeauRasio07,Kremer+20,Rodriguez+22}. Given these uncertainties, we test different treatments of the mass loss and collision outcomes, documented in Section~\ref{sec:massloss} and compare their results in Section~\ref{sec:FMLcomparison}. Section~\ref{sec:relax} and \ref{sec:Demographics} present and discuss our findings in the context of the stellar demographics and mass function. Sections~\ref{sec:Gobjects} and Section~\ref{sec:RGs} discusses implications for the G object and RG populations, respectively. In Section~\ref{sec:discussion}, we summarize our findings.

\begin{figure*}
	\includegraphics[width=0.95\textwidth]{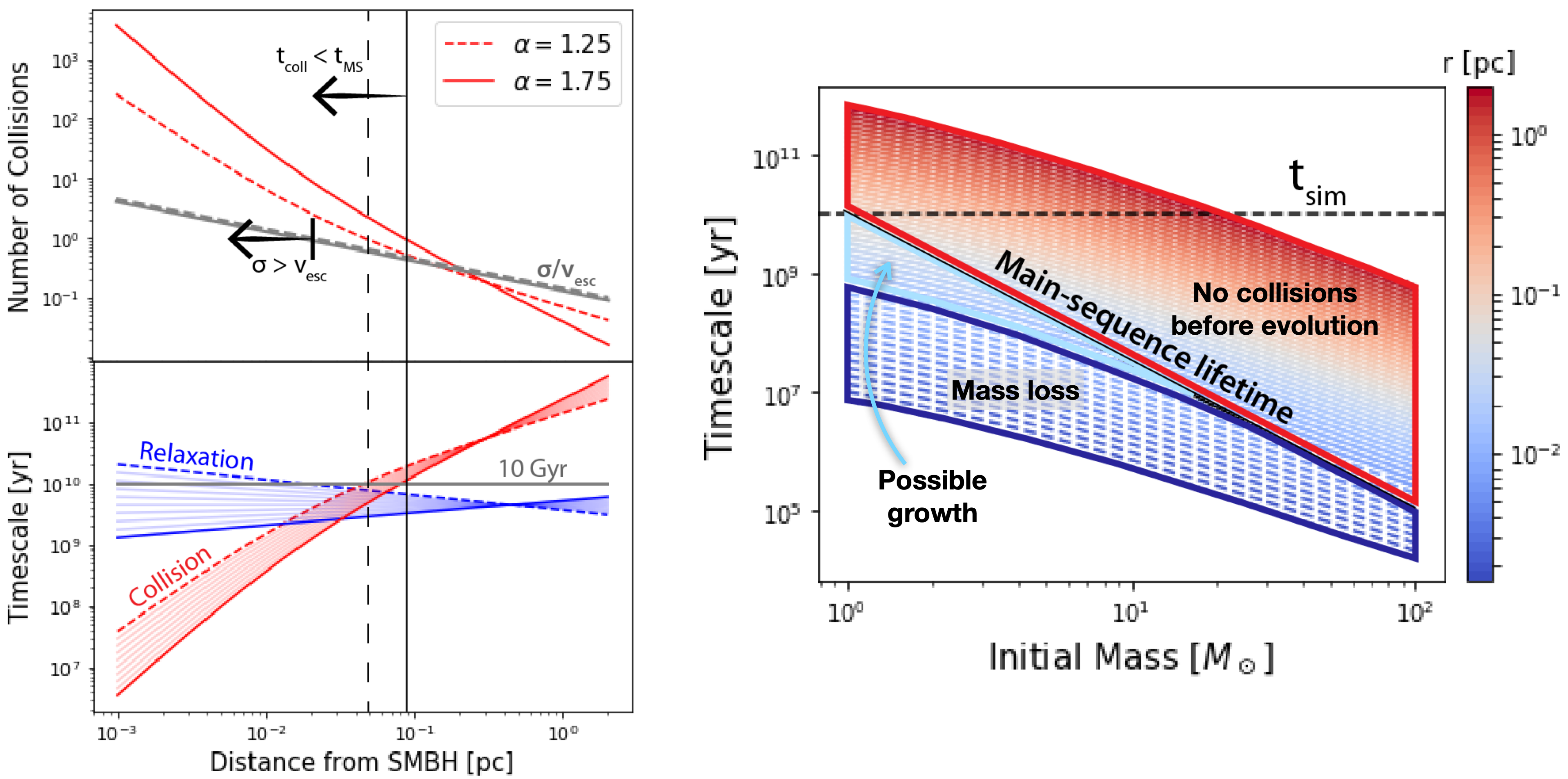}
	\caption{ \textbf{Left lower panel:} We plot the relaxation (blue) and collision (red) timescales for a range of stellar density profiles, from $\alpha = 1.25$ (dashed) to $1.75$ (solid). To guide the eye, we plot the approximate lifetime, $10$~Gyr, of a solar mass star in grey. The dashed and solid lines indicate the distances from the SMBH at which the star's lifetime is equivalent to the collision timescale for the two stellar density profiles considered in this study. 
	\textbf{Left upper panel:} We plot the approximate number of collisions, which is given by the main-sequence lifetime $10$~Gyr divided by the collision timescale, as a function of distance from the SMBH. When this number is greater than one, it should be taken as an indicator of the importance of collisions, not the exact number of collisions that will occur, because this estimate assumes that the star's cross-section and mass remain unchanged following each collision. When this number is less than one, it is equivalent to the probability that a solar mass star will experience a collision over its lifetime.
    In grey, we also plot the velocity dispersion divided by the escape speed from a solar mass star. Where $\sigma > v_\mathrm{esc}$, collisions may result in significant mass loss.
    \textbf{Right panel:} Similar the lefthand plot, here we compare the collision timescale to other key timescales in our simulation. We calculate the collision timescale as a function of stellar mass and, by extension, stellar radius using Eq.~(\ref{eq:t_coll_main_ecc}). We plot this timescale for different distances from the SMBH, as indicated by the colorbar. The dashed black line represents maximum simulation time, $10$~Gyr, while the solid black line shows the approximate main-sequence lifetime as a function of mass. Wherever the collision timescale is greater than these two timescales, the chances of a star experiencing a collision, based on its mass and location, are very low. Wherever the collision timescales is shorter than $t_\mathrm{MS}$, there is a high probability that a star will undergo one or more collisions over its lifetime. Dashed lines represent cases when the velocity dispersion exceeds the escape velocity from the star, a tracer for where collisions tend to be destructive. We have highlighted this region in dark blue.} 
     \label{fig:timescales}
\end{figure*}

\section{Method} \label{sec:methodology}

\subsection{Initial Conditions}

We follow a sample population of $1000$ stars embedded in the nuclear star cluster. We draw the initial masses of our sample population using a Kroupa initial mass function (IMF). We limit the stars to masses between $0.5$ and $100$~M$_\odot$. However, in practice, the most massive stars in our simulations tend to be around $30$~M$_\odot$ because of the steep IMF. We draw the orbital eccentricities of the stars from a thermal distribution. Their semimajor axes are drawn from a uniform distribution in log distance, i.e., one in which $dN/d(\log r)$ is constant. This distribution ensures adequate sampling at all distances from the SMBH, necessary to build comprehensive understanding of stellar collisions in the GC.

Each star in our sample population is 
influenced by the surrounding stellar cluster, which we treat as a reservoir. For simplicity, we assume that the surrounding stars have an average mass of $1$~M$_\odot$, which remains constant over the duration of our simulation. We describe the density of the cluster as a function of $r_\bullet$, the distance from the SMBH, using a power law with the form:
\begin{eqnarray} \label{eq:density}
    \rho(r_\bullet) = \rho_0 \left( \frac{r_\bullet}{r_0}\right)^{-\alpha} \ , 
\end{eqnarray}
where $\alpha$ determines the slope of the density profile.
For the inner parsec of the GC, the density is normalized using the observationally-constrained values $\rho_0 = 1.35 \times 10^6 \, M_\odot/{\rm pc}^3$ at $r_0 = 0.25 \, {\rm pc}$ \citep{Genzel+10}. For a uniform population of solar mass stars, the number density becomes:
\begin{eqnarray}
    n(r_\bullet) = \frac {\rho(r_\bullet)}{1 \, M_\odot} \ . \label{eq:density_stars}
\end{eqnarray}
Unless otherwise noted, we assume that the surrounding star cluster is dynamically relaxed, resulting in a cuspy Bahcall-Wolf profile 
\citep[i.e., $\alpha = 1.75$][]{BahcallWolf76,Bar-Or+13,AlexanderHopman+09,Keshet+09,AharonPerets16}. The true density profile in the GC may be shallower, with an index closer to $1.5$ or $1.25$ \citep[e.g.,][]{Buchholz+09,Do+09,Bartko+10,Gallego-Cano+18,Gallego+20,Schodel+14,Schodel+18,Schodel+20,LinialSari22}.
Appendix~\ref{sec:moreresults} shows results that use $\alpha = 1.25$. The most realistic picture of the GC may lie somewhere between the $\alpha = 1.25$ and $1.75$ results.

The mass of the SMBH $M_{\bullet}$ sets the velocity dispersion of the cluster:
\begin{eqnarray}\label{eq:sigma}
    \sigma(r_\bullet) = \sqrt{ \frac{GM_{\bullet}}{r_\bullet(1+\alpha)}},
\end{eqnarray}
where $\alpha$ is the slope of the density profile \citep{Alexander99,AlexanderPfuhl14}. $\sigma$ is approximately the orbital velocity. Together, the velocity dispersion and number density govern key physical processes like collisions because these properties determine the frequency of interactions between a star and its surrounding objects.

\subsection{Direct Collisions} \label{sec:collision}

A star in the GC is expected to experience a collision over a characteristic timescale, $t_{\rm coll}$. This timescale can be estimated using an $(n \sigma A)^{-1}$ rate calculation, where $A$ is the cross-section of interaction and $n$ and $\sigma$ are determined by Eqs~(\ref{eq:density}) and (\ref{eq:sigma}). The collision timescale also has a weak dependence on eccentricity:
\begin{eqnarray} \label{eq:t_coll_main_ecc}
     t_{\rm coll}^{-1} &=& \pi n(a_\bullet) \sigma(a_\bullet) \nonumber \\ &\times& \left(f_1(e_\bullet)r_c^2 + f_2(e_\bullet)r_c \frac{2G(M_\odot+M)}{\sigma(a_\bullet)^2}\right)\ .
\end{eqnarray}
where $f_1(e_\bullet)$ and $f_2(e_\bullet)$ are given by \citet{Rose+20}, $G$ is the gravitational constant, and $r_c$ is the sum of the radii of the colliding stars.

We plot the collision timescale for different density profiles, spanning $\alpha = 1.25$ (dashed) to $\alpha = 1.75$ (solid), in red in the left bottom panel of Figure~\ref{fig:timescales}. In the left upper panel, we plot the number of collisions expected per star as a function of distance from the SMBH. The number of collisions per star are also plotted in red, with dashed and solid lines corresponding to the $\alpha = 1.25$ and $\alpha = 1.75$ density profiles, respectively. These curves assume a uniform population of solar mass stars. On the plot, we use grey vertical lines to mark the distance within which each star will experience at least one collision. This distance corresponds to where the collision timescale is equivalent to the main-sequence lifetime of a solar mass star. Outside of this distance, the number of collisions per star is less than one, indicating the probability that a single star will experience a collision or, equivalently, the fraction of the population that collides. In the upper panel, we also show the velocity dispersion as a function of distance from the SMBH in units of escape velocity from the star. Since we take the velocity dispersion to be the relative velocity of the two stars as they collide, this grey curve is a measure of how energetic each collision is. Where $\sigma$ is greater than the escape speed from a solar mass star, we expect the collisions to be destructive, leading to mass loss. We discuss collision outcomes in greater detail in the next section.

In our simulations, we account for collisions using a statistical approach following \citet{Rose+22}. Over a timestep $\Delta t$, the probability that a given BH will experience a collision is approximately $\Delta t/t_\mathrm{coll}$. We draw a random number between 0 and 1. If that number is less than or equal to the probability, we assume a collision has occurred. We adjust the mass of the star accordingly and repeat this process until the simulation has reached the desired total time, $10$~Gyr. As in \citet{Rose+22}, we use a timestep of $1 \times 10^6$~yr.


\subsection{Collision Outcomes} \label{sec:massloss}
\subsubsection{General Overview}
The outcome of a direct collision between two stars depends on many factors, including the velocity dispersion, the impact parameter, and assumptions about the stellar structure \citep[e.g.,][]{Lai+93,FreitagBenz}. Given the uncertain outcome of stellar collisions, especially at high velocities, we explore a range of possibilities, varying our treatment of the mass loss and merger products. Broadly, there are two possible outcomes 
of a collision.
\begin{enumerate}
    \item \textbf{Merger:} In this case, the star in question gains a solar mass. We reiterate that in our toy model, the surrounding cluster contains only $1$ $\mathrm{M}_{\odot}$ stars, while the masses of our sample are drawn from a Kroupa IMF. From the merger product, some mass may be ejected depending on the nature of the collision. We estimate the final mass
    $M_f$ of the star as
    $M_f = M_i + 1 -f_{ml}(M_i+1)$, or $M_f = (1 -f_{ml})(M_i+1)$, where $M_i$ is the stellar mass before the collision took place and $f_{ml}$ is the fractional mass loss caused by the collision, discussed in greater detail below.
    \item \textbf{Mass Loss Only:} In the second case, there is no merger -- the primary star fails to capture the collider -- but the star still loses some fraction of its mass following the impact. Its final mass equals $(1-f_{ml})M_i$.
\end{enumerate}

In the following subsections, we detail different physically motivated approaches to estimating the fractional mass loss. Following a collision, we recalculate the radius of the star using its updated mass. The thermal timescale is short enough ($10^7$ yr) for us to take the radius of a main-sequence star, except for the low mass ($< 1$ $\mathrm{M}_\odot$) stars. For these low-mass stars, the thermal timescale may be longer than the collision time, meaning they remain distended when the next collision occurs. The larger radius than accounted for in our model may hasten their destruction.

Irrespective of the specific prescription adopted for $f_{ml}$, we can make predictions about where in the GC certain outcomes are likely to occur. Figure \ref{fig:timescales}, right panel, illustrates the parameter space in which different outcomes are expected. In particular, the right plot shows the collision timescale as a function of the star's mass at different distances from the SMBH. Each collision timescale curve is color-coded according to distance from the SMBH as indicated by the color bar on the left side of the plot. We also include the main-sequence lifetime as a function of stellar mass (solid black line) and the simulation time of $10$~Gyr (dashed black line). The region highlighted in red represents a regime in which most stars will evolve off the main-sequence before a collision can occur. On the other hand, when the collision timescale is less than the main-sequence lifetime, stars may experience one or more collisions before evolving off the main sequence. The regime highlighted in dark blue in the right hand plot represents the scenario where the 
dispersion velocity on the onset of the collision is larger than the escape velocity of the two colliding stars (i.e., $\sigma > v_{\rm esc}$). Thus, we expect collisions to result in significant mass loss from the stars and no mergers to occur. However, in the small light blue wedge, the dispersion velocity is smaller than the escape velocity, and collisions may result in mergers and mass growth. As mentioned above, the details of the mass loss and growth are uncertain, and we consider several scenarios below.

\begin{figure}
	\includegraphics[width=0.99\columnwidth]{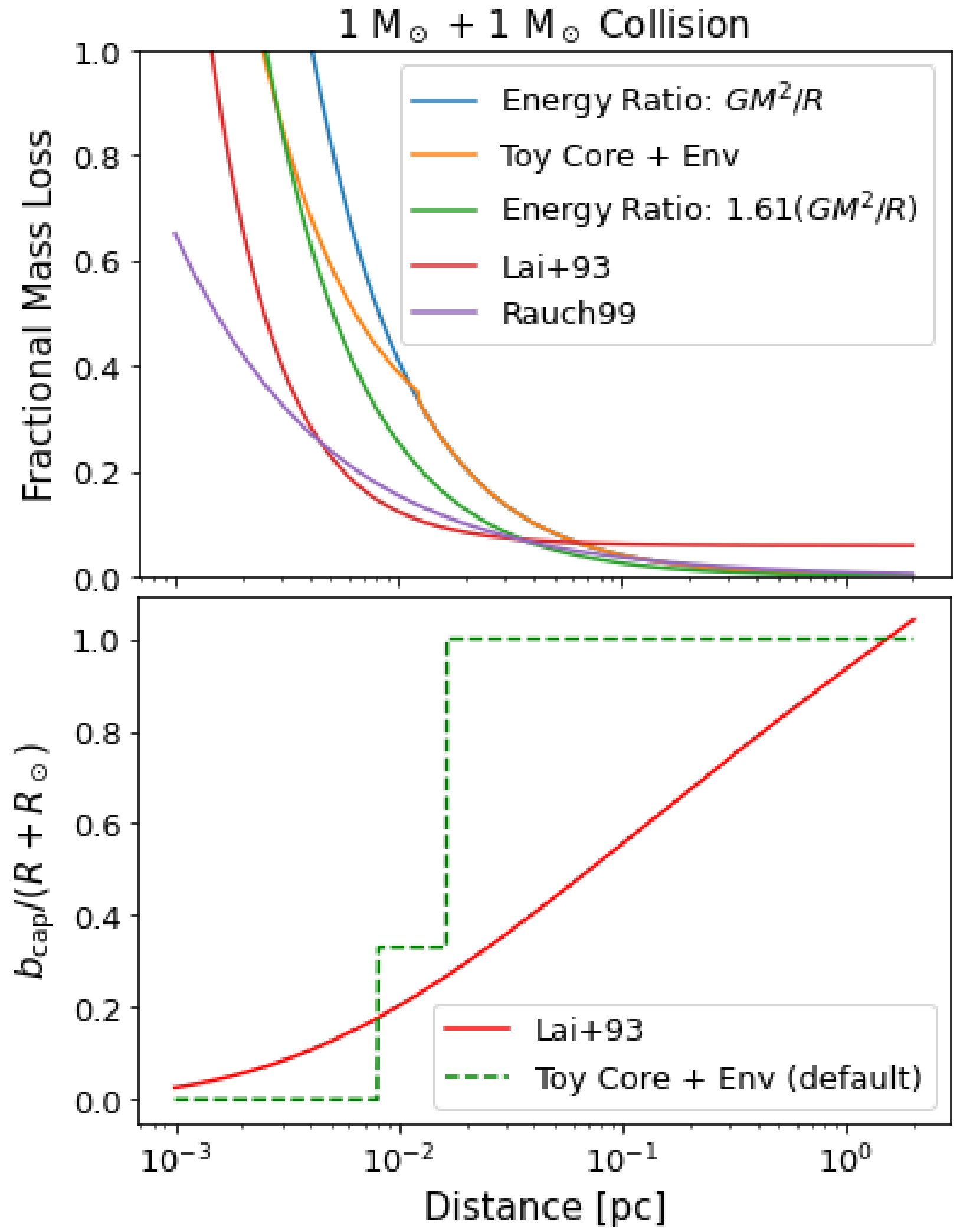}
	\caption{\textbf{Upper Panel:} The fractional mass loss expected from a head-on collision between two solar mass stars is plotted as a function of distance from the SMBH for different models. This is the maximum fractional mass loss that can be produced by a collision between Sun-like stars; an impact parameter larger than zero can only result in less mass loss. The blue line estimates the $f_{ml}$ using Eq.~\ref{eq:fml_heur}, where the binding energy of the stars is approximated as $GM^2/R$. The green curve uses a similar estimate, but includes a prefactor in the binding energy to account for the fact that the stellar density increases near the core. The orange curve combines these two estimates to construct a toy model of the star: it has a loosely bound ``envelope'' containing about $30$ \% of the star's mass and a more tightly bound ``core.'' The red and purple curves use the fitting formulae from \citet{Lai+93} and \citet{Rauch99}, respectively. \textbf{Lower Panel:} Here we show the two different prescriptions we use to determine whether a merger occurs following a collision or, in other words, whether the stars manage to capture each other after impact. The red curve is based on the fitting formulae from \citet{Lai+93}, while the green, dashed curve shows the toy model approach taken in most of our simulations. The latter model compares the escape speed from the ``core'' and surface of the star to the velocity dispersion in order to determine whether a merger occurs. This prescription is used in all simulations except those specifically labeled as ``Lai+93''.
 }
     \label{fig:FML}
\end{figure}

\subsubsection{Energy Ratio Estimate of the Mass Loss} \label{sec:heuristic}

The fractional mass loss can be approximated as the ratio of the kinetic energy to the binding energy of the two colliding stars:
\begin{eqnarray}
    f_{ml} = \frac{\mu \sigma^2}{GM_i^2/R_i+GM_\odot^2/R_\odot}, \label{eq:fml_heur}
\end{eqnarray}
where $\mu$ is the reduced mass of the two colliding stars and $M_i$ and $R_i$ are the mass and radius of the star from our sample population.
This equation is roughly consistent with equations presented in the qualitative overview of collisions in \citet{Lai+93}, with some adjustments to make the fractional mass-loss unitless. The aforementioned study includes a parameter $\beta$ in the denominator of $f_{ml}$,  a variable that accounts for the gas pressure in the star. While $\beta$ is order unity for solar mass stars, radiation pressure becomes important for more massive stars; it is easier to unbind material from them during a collision \citep{Lai+93}. The main-sequence lifetime for these massive stars, however, is short compared to the collision timescale. Their low probability of experiencing a collision means that our assumption that $\beta = 1$ does not significantly impact our results.

Stars are not uniform density spheres; in fact, most of a star's mass is concentrated near its center, leading to a higher binding energy than the estimate in the denominator of Eq.~\ref{eq:fml_heur} \citep[e.g.,][]{Christensen-Dalsgaard+96}. To illustrate this point, we plot of the mass enclosed as a function of radius within the Sun based on \citet{C-Dalsgaard+96} in Figure~\ref{fig:SSM} (see Appendix~\ref{sec:FML}). A spherical object with this mass profile has binding energy $1.61 G M^2/R$. For a slightly more accurate estimate of the fractional mass loss, we include this prefactor, $1.61$, in Eq.~\ref{eq:fml_heur}. In Figure~\ref{fig:FML}, we plot the fractional mass loss with and without the prefactor in green and blue, respectively, for a collision between two $1$~M$_\odot$ stars. The fractional mass loss decreases with distance from the SMBH because it depends on the velocity dispersion.

In this prescription, if $\sigma < \sqrt{2GM/R}$, roughly the escape velocity from the stars, we assume that a merger occurs (see Figure \ref{fig:timescales}). Otherwise, we calculate the final mass of the star as $M_\mathrm{post} = M(1-f_{ml})$. This prescription over-predicts the amount of mass lost for reasons discussed in Section~\ref{sec:Lai93}.

\subsubsection{Toy Core+Envelope Model} \label{sec:toycore}
The purpose of this toy model is to provide physical insight into the outcome of stellar collisions. Most of a star's mass is concentrated at its center. As a result, the escape velocity from the surface of the star may be much lower than that from the core, a subtlety not fully encapsulated by the approximations in Section~\ref{sec:heuristic}. Here we construct a more nuanced toy model based on the internal structure of the Sun \citep[e.g.,][]{C-Dalsgaard+96}. Based on the mass enclosed, $M_\mathrm{enc}$, within a radius $r$ in the Sun, we calculate the escape velocity from a sphere with equivalent mass and radius, $v_\mathrm{esc} = \sqrt{2GM_\mathrm{enc}/r}$, as a function of $r$. We determine that this $v_\mathrm{esc}$ reaches a maximum of about $800$~$\mathrm{km/s}$ at $R = 0.33 R_\odot$ (see Figure~\ref{fig:SSM} and Appendix~\ref{sec:FML}). This radius encloses almost exactly two thirds of the Sun's mass. We define this region as the ``core,'' though we stress that it is not representative of the true core of the Sun; in our simple model, there is a tightly bound inner part of the star and a more loosely bound outer ``envelope.''

We calculate the fractional mass loss as follows. We begin by calculating the ratio of the kinetic energy to the binding energy of the stars as in Eq.~(\ref{eq:fml_heur}). We then take the maximum between the corresponding change in mass and the mass of the ``envelope,'' $0.33$~M$_\odot$. If the kinetic energy is greater than the binding energy of the envelope, we assume the remaining kinetic energy can go into ejecting material from the more tightly bound ``core.'' In other words, if the collision has enough energy to unbind more than the outer third of the star's mass, the total mass lost equals:
\begin{eqnarray}
    M_\mathrm{lost} = M_\mathrm{env}+\frac{\mu \sigma^2/2-U_\mathrm{env}}{GM_\mathrm{core}^2/R_\mathrm{core}} \ , \label{eq:fml_heur}
\end{eqnarray}
for a collision between two $1$~M$_\odot$ stars. As noted above, in this equation $M_\mathrm{core}$ is about $0.66$~M$_\odot$ and $R_\mathrm{core}$ is about $0.33 R_\odot$. In Figure~\ref{fig:FML}, we plot the fractional mass loss in orange.

The main advantage of this prescription is that it can be expanded to account for an impact parameter. From a geometric standpoint, head-on collisions are unlikely. Collisions with a non-zero impact parameter result in less mass loss because only the outer envelopes of the stars interact, while most of the mass is concentrated in the core. Accounting for the impact parameter, $b$ is therefore crucial in this work. We draw an impact parameter between zero (a head-on collision) and the sum of the two stellar radii in question, $R+R_\odot$ (a grazing collision) using a probability density that goes as $b^2/(R+R_\odot)^2$.

In our simple model, during a collision with $R_\mathrm{core} < b$, only the envelopes interact, and there cannot be any additional mass loss from the core. On the other hand, any collision with an impact parameter smaller than the core radius can result in mass loss from this region. Additionally, in this prescription, there are two sets of conditions that, when met, result in a merger: $b<2R_\mathrm{core}$ and $\sigma<v_\mathrm{esc,core} \sim 800$~$\mathrm{km/s}$, or $2R_\mathrm{core}<b<2R_\odot$ and $\sigma<v_\mathrm{esc,surface} \sim 600$~$\mathrm{km/s}$. Note that this formula for drawing the impact parameter does not account for gravitational focusing, which can make head-on collisions more likely and lead to slightly higher fractional mass loss.

In order to expand this approach to stars of different masses, we assume that all stars have a similar structure to our Sun, allowing us to scale the quantities described above. In other words, we calculate the binding energy of the star as $1.61 G M^2/R$, and assume that about two thirds of the star's mass is concentrated in the inner one third of the radius. While the true structure of a star depends on its mass, which determines the convective regions, stars with $M \gtrsim 1$~M$_\odot$ have a similar structure. Massive stars ($M \gtrsim 10$~M$_\odot$) have convective cores. However, their shorter main-sequence lifetime coupled with their low abundances in our sample means that very few of them collide. Future work may consider a more accurate approach to stars of other masses, perhaps by using different polytropic indices as was done by \citet{RubinLoeb}, but for the purposes of this study, the approach detailed above provides a simple mass-loss recipe that complements the other prescriptions used in this study (see Figure~\ref{fig:FML}).


\subsubsection{Fitting Formulae}  \label{sec:Lai93}
In other simulations, we rely on fitting formulae from on smooth-particle hydrodynamics (SPH) studies of collisions at high velocities. Considering collisions in galactic nuclei, \citet{Lai+93} explores a wide variety of relative velocities, mass ratios, and impact parameters. This study provides a variety of fitting formulae, physically reasoned and based on their numerical results, that calculate the fractional mass-loss and determine whether or not a collision results in a merger. Specifically, we use their 
equations
(4.4) and (4.8)-(4.10) and the fitting parameters in their Table 1. We label simulations that use these fitting formulae as ``Lai+93''.

As an alternative, we also use the fitting formulae in Appendix B of \citet{Rauch99}. Based on a set of SPH simulations for high impact velocities by M. Davies, equation B1 relates the fractional mass loss of the two stars in question using their mass ratio, and equation B2 calculates the total fractional mass loss from the system in terms of the impact parameter, mass ratio, and relative velocity. The label ``Rauch99'' denotes simulations in this paper that use these equations.

In Figure~\ref{fig:FML}, we plot the fractional mass loss expected from these fitting formulae for a head-on collision ($b = 0$) between two $1$~M$_\odot$ stars as a function of distance from the SMBH. The equations from \citet{Rauch99} generally give a lower fractional mass loss than the other models used in this work.\footnote{For additional comparisons of these mass-loss predictions with other studies, please see figures~5 and 6 in \citet{RubinLoeb} and figures~12 and 13 in \citet{FreitagBenz}}. We note that both \citet{Rauch99} and \citet{Lai+93} assume that stars are polytropes, which may provide an incomplete picture of the full range of collision outcomes \citep{FreitagBenz}. Additionally, \citet{Rauch99} note the difficulties in determining the post-collision radius of the star and approximate it as being equal to that of a main-sequence star with the same mass, an approach that is consistent with ours.

In the bottom panel of Figure~\ref{fig:FML}, we also plot the maximum impact parameter, $b_\mathrm{max}$, that will result in a merger as a function of distance from the SMBH, which acts as a proxy for the velocity dispersion. The red line uses the fitting formala provided in \citet{Lai+93}, while the green dashed curve represents our simple toy model approach, used as the default in all of our other simulations. Unlike \citet{Lai+93}, whose equations give the specific impact parameter needed to achieve a merger at a given impact velocity, \citet{Rauch99} use an escape velocity argument to determine whether or not a merger occurs, much like our own toy model. We therefore adopt the conditions described in Section~\ref{sec:toycore} (green dashed line in the bottom panel of Figure~\ref{fig:FML}) to determine whether or not a collision results in a merger in Rauch99 simulations. These conditions lead to more mergers overall than the \citet{Lai+93} fitting formula. However, as noted in Section~\ref{sec:toycore}, our formula for drawing the impact parameter does not account for gravitational focusing, which makes head-on collisions more likely. Our Lai+93 simulations may therefore underestimate the number of mergers because of the stricter $b_\mathrm{max}$ condition. Our Rauch99 simulations, which use the less restrictive $b_\mathrm{max}$ merger conditions, may provide a more accurate picture of mergers outside of $0.02$~pc, where gravitational focusing becomes increasingly important.

\subsection{Two-Body Relaxation}
As a star orbits the SMBH, it will experience frequent weak gravitational interactions with passing objects. Over time, the small effects of these interactions accumulate, altering the star’s orbit. This process is known as relaxation and acts over an associated timescale, $t_\mathrm{rlx}$, the amount of time needed for the star's orbital energy to change by order of itself \citep[e.g.,][]{BinneyTremaine}:
\begin{eqnarray} \label{eq:t_rlx}
t_{\rm rlx} = 0.34 \frac{\sigma^3}{G^2 \rho \langle M_\ast \rangle \ln \Lambda_{\rm rlx}} \ ,
\end{eqnarray}
In Figure~\ref{fig:timescales}, the blue lines show this timescale as a function of distance from the SMBH for a range of stellar density profiles.

We simulate the effects of relaxation in our code. Over each timestep, we apply a small instantaneous velocity kick to the star, from which we calculate the new, slightly altered orbital parameters \citep[see][for the full set of equations]{Lu+19,Naoz+22}. The velocity kick is drawn from a Gaussian distribution with a standard deviation $\Delta v_{rlx}/\sqrt{3}$, where $ \Delta v_{rlx} = v_\mathrm{orbital} \sqrt{ {P_\mathrm{orbital}}/{t_{rlx}}}$ \citep[e.g.,][]{Bradnick+17,Rose+22,Naoz+22}. Here, $v_\mathrm{orbital}$ and $P_\mathrm{orbital}$ are the orbital speed and period of the star in question. This prescription allows us to account for changes in the orbital parameters over time from interactions with the surrounding stars.

\subsection{Stopping Conditions}

At each timestep in our code, we calculate the star's main-sequence lifetime, $t_\mathrm{MS}$, according to its present mass: $t_\mathrm{MS} = 10^{10} (M/M_\odot)^{-2.5}$. When a star's age exceeds its main-sequence lifetime, we terminate the simulation. In the case of a merger, we calculate the new age of the merged product using a ``sticky-sphere model'' \citep[see, e.g.,][]{Kremer+20}. Mixing during the collision and merger can introduce a fresh supply of hydrogen to the core, extending the lifetime of the merged star. The dimensionless parameter $f_\mathrm{rej}$, ranging from $0.1$ to $1$, quantifies the degree of mixing and therefore the degree of rejuvenation \citep[e.g.,][]{Hurley02,Breivik+20}. While the true value of $f_\mathrm{rej}$ is challenging to ascertain since it can depend on the details of each collision and stellar structure, we adopt the $f_\mathrm{rej} = 1$, the most conservative value. The new age of the star, $t_\mathrm{age,f}$, can be calculated as follows:
\begin{eqnarray}
   t_\mathrm{age,f} = f_\mathrm{rej} \frac{t_\mathrm{MS,M+M_\odot}}{M+M_\odot} \left( \frac{M \times t_\mathrm{age,i}}{t_\mathrm{MS,M}} + \frac{M_\odot \times t_\mathrm{sim}}{t_\mathrm{MS,\odot}} \right). \label{eq:stickylifetime}
\end{eqnarray}
Above, $t_\mathrm{age,i}$ is the age of the star in question before it collides with a solar mass star -- recall that we assume that solar mass stars comprise the surrounding cluster -- and $t_\mathrm{MS,M}$ is its main-sequence lifetime. The star with which it collides has main-sequence lifetime $t_\mathrm{MS,\odot}$ and age equal to the time ellapsed in the simulation, $t_\mathrm{sim}$. $t_\mathrm{MS,M+M_\odot-M_\mathrm{lost}}$ denotes the main-sequence lifetime of the merger product. 


In addition to the stellar age, we include two other stopping conditions in our code. Relaxation processes may drive a star's orbital eccentricity to high values, shrinking its pericenter passage to fall within the tidal radius of the SMBH. We include a stopping condition to account for these tidally disrupted stars:
\begin{eqnarray}
    a(1-e) < R(2 M_\bullet/M)^{1/3}. \label{eq:TDE}
\end{eqnarray}
Lastly, some stars undergo significant mass loss through one or more collisions. The code terminates once the mass of a star falls below $0.1$~M$_\odot$.


\begin{figure}
	\includegraphics[width=0.97\columnwidth]{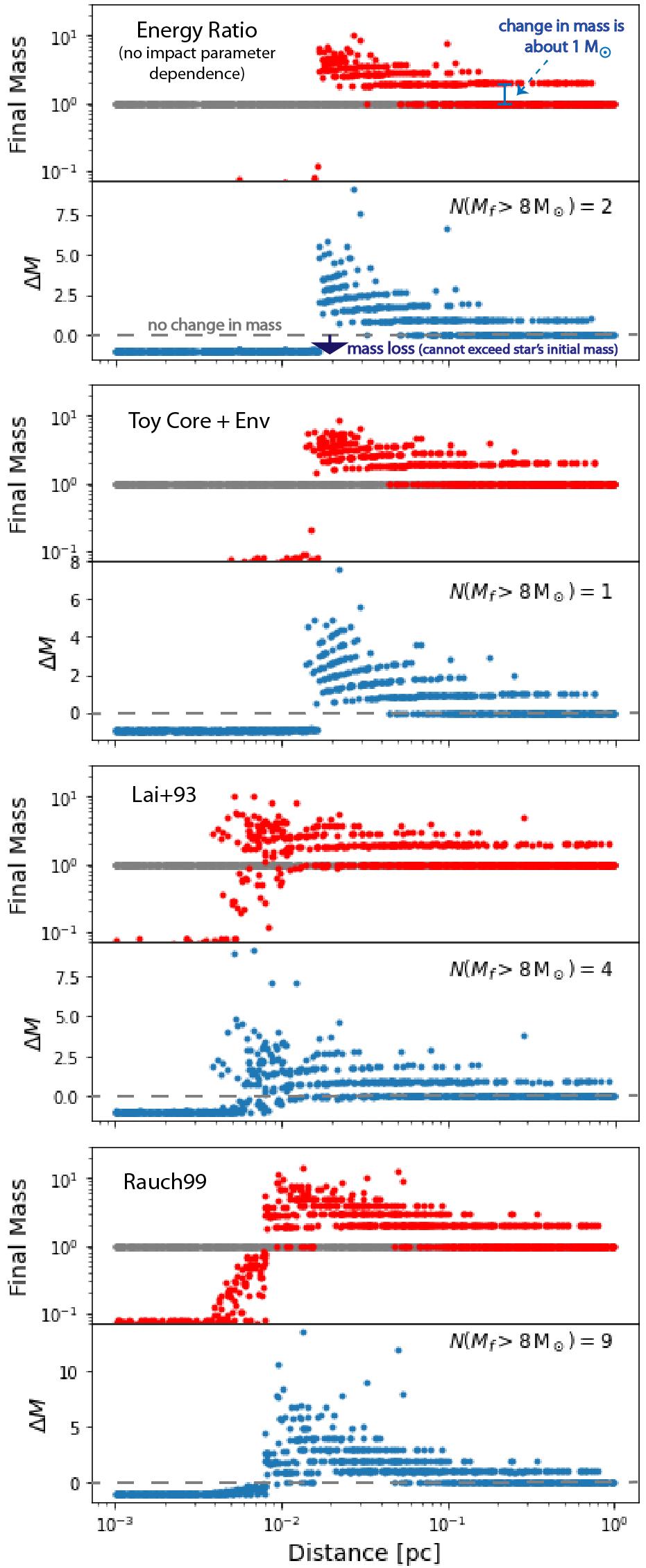}
	\caption{We compare simulated results for different fractional mass loss recipes. So that the differences are easy to discern, we do not include relaxation in these simulations. We begin with a uniform population of $1$~M$_\odot$ stars distributed uniformly in log distance from the SMBH. Grey dots represent the initial masses and distances. From top to bottom, we show simulation results using an energy ratio estimate for the fractional mass loss (Section~\ref{sec:heuristic}), a toy core plus envelope model for the stars (Section~\ref{sec:toycore}), and fitting formulae from comprehensive studies of collisions using SPH simulations (Section~\ref{sec:Lai93}). The final masses are shown in red in the upper panels, and the blue dots in the lower panels represent the change in mass of each star.}
     \label{fig:SMM}
\end{figure}

\section{Comparing Mass Loss Prescriptions: Low Mass Remnants and Massive Merger Products} \label{sec:FMLcomparison}

We begin by comparing the different mass loss prescriptions detailed in Sections~\ref{sec:heuristic}, \ref{sec:toycore}, and \ref{sec:Lai93}. We therefore 
run several simulations with a sample population of only $1$~M$_\odot$ stars and identical initial conditions, constraints we later relax in Section \ref{sec:Demographics}. At this stage, we do not include relaxation in order to build a clear picture of the differences between our mass loss prescriptions. Figure \ref{fig:SMM} juxtaposes the results. The initial conditions, $1000$ solar mass stars distributed uniformly in log distance from the SMBH, are shown in grey, while red dots show the final masses versus distance from the SMBH. The blue dots show the total change in each star's mass over the simulation. Each row corresponds to a different mass loss prescription, indicated by the label in the left corner.

All of the mass loss prescriptions yield qualitatively similar results, consistent with the expectations from Figure~\ref{fig:timescales}. Close to the SMBH, collisions work to destroy the stars, while $\sim 1$~pc from the SMBH few collisions occur. In between these two extremes, one or more collisions occur, and they result in mergers. Generally, our results suggest that multiple collisions between $1$~M$_\odot$ stars can produce more massive stars.
These massive products are localized between $0.01$ and $0.1$ pc from the SMBH. In this region, a few conditions must align to support the formation of massive stars. Firstly, the collision timescale is shorter, in some cases by an order of magnitude, than the main-sequence lifetime and several collisions can take place. Secondly, the velocity dispersion is small enough that stars can capture each other and the factional mass loss is low. The final mass of the stars formed through collisions therefore depends on the specific mass loss prescription adopted.

For these initial conditions, all of our mass loss prescriptions can produce a $>8$~M$_\odot$ star. However, the Rauch99 prescription was the only one to do so consistently. This outcome is not surprising given that the Rauch99 equations result in less mass loss compared to the others (see Figure~\ref{fig:FML}), producing more massive stars.
Over several Rauch99 simulations with $1000$ solar mass stars each, we find that a few stars out of the $1000$ will exceed $10$~M$_\odot$ through mergers with other low-mass stars. We scale this number using a realistic density profile normalized by the $M$-$\sigma$ relation \citep{BahcallWolf76,Tremaine+02}. Accounting for the number of stars residing between $0.01$ and $0.1$ pc in the Galactic Center, we estimate that there may be more than $\sim 100$ stars over $10$~M$_\odot$ formed through collisions between lower mass stars. This estimate assumes that collisions tend to produce less mass loss, as in the Rauch99 prescription, and result in only moderate rejuvenation ($f_\mathrm{rej} = 1$ in Eq.~\ref{eq:stickylifetime}). If mergers in the Galactic Center are more effective at rejuvenating stars, then collisional processes should lead to even more massive stars and in greater abundance.

Another difference between mass loss prescriptions is their ability to produce low mass remnants. Often, a collision within $\sim 0.01$~pc of the SMBH, where the velocity dispersion is high, will only remove the outer layers of a star. This outcome is common because most collisions have a non-zero impact parameter \cite[see also discussion in][]{Rauch99}. The contrast between the upper row and last two rows in Figure~\ref{fig:SMM} illustrate the importance of the impact parameter. The upper row does not include an impact parameter dependence, and as a result it overestimates the number of stars that are destroyed. Rauch99, on the other hand, predicts many low-mass collision remnants, a result discussed in their original study, \citet{Rauch99}. The evolution of these remnants post-collision is uncertain. However, from a dynamical standpoint, these objects may be long-lived. Their small cross-section leads to a longer collision timescale, making it less likely that they will undergo another destructive encounter.


\section{The Effect of Relaxation} \label{sec:relax}
\begin{figure}
	\includegraphics[width=0.95\columnwidth]{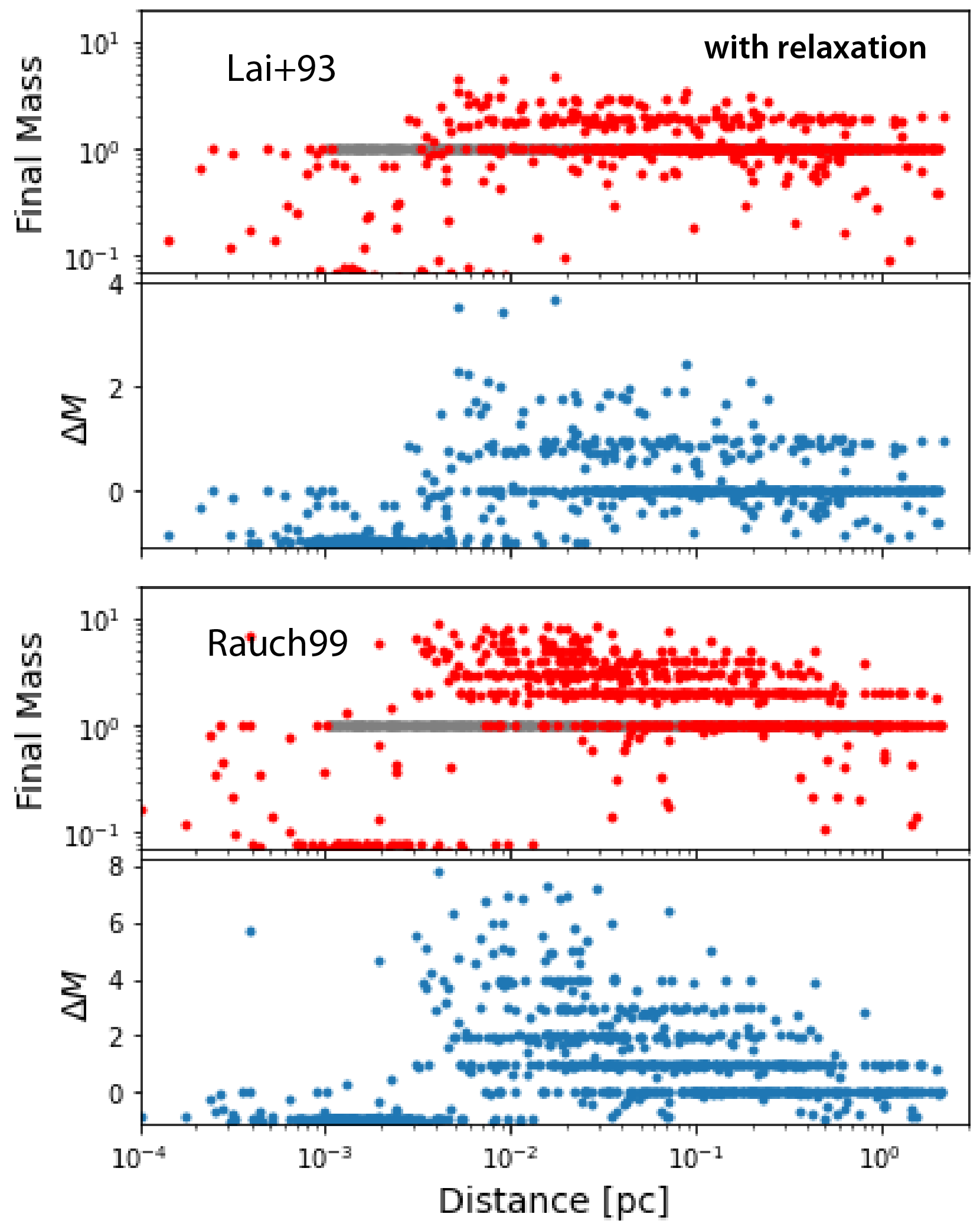}
	\caption{The same as the third and fourth row of Figure~\ref{fig:SMM}, but with the addition of relaxation. We begin with a uniform stellar population of solar mass stars distributed uniformly in log distance from the SMBH.}
     \label{fig:SMMrlx}
\end{figure}

In the previous section, we demonstrate that the general collision outcome trends described in Figure~\ref{fig:timescales} hold. Relaxation has the potential to complicate the physical picture, however, by allowing stars to diffuse into different regions from the SMBH. We show the effects of this dynamical process in Figure~\ref{fig:SMMrlx}. The simulations shown in the Figure use identical initial conditions as those in last two rows of Figure~\ref{fig:SMM}, except with the addition of relaxation. One main difference is that due to two-body relaxation, low mass remnants can migrate out of the region in which they form, around $0.01$~pc, to larger distances from the SMBH. We therefore predict a diffuse population of peculiar low-mass stars throughout the GC. In contrast, stars that experience the most mass growth are still localized to $0.01$-$0.1$~pc from the SMBH. These stars must form in this region because otherwise either the collision timescale is too long to achieve multiple mergers, or the collisions are destructive in nature. Additionally, as a star grows in mass, its main-sequence lifetime becomes short compared to the relaxation timescale (Figure~\ref{fig:timescales}). As a result, the most massive stars have insufficient time to diffuse out of the region in which they form. We therefore expect massive stars formed through stellar collisions to be limited within $ \lesssim 0.1$~pc from the SMBH.

\section{Stellar Demographics}\label{sec:Demographics}

Thus far, we have considered only uniform populations of $1$~M$_\odot$ stars. In this section, we use a Kroupa IMF to generate our initial conditions and explore general trends in the stellar mass distributions. We consider two mass-loss prescriptions, Lai+93 and Rauch99, because they encapsulate the range of results in Figure~\ref{fig:SMM}. Recall that  Rauch99 favors mergers with minimal mass loss, while the Lai+93 equations lead to greater mass loss and have a stricter merger condition.

\begin{figure*}
\begin{center}
	\includegraphics[width=0.99\textwidth]{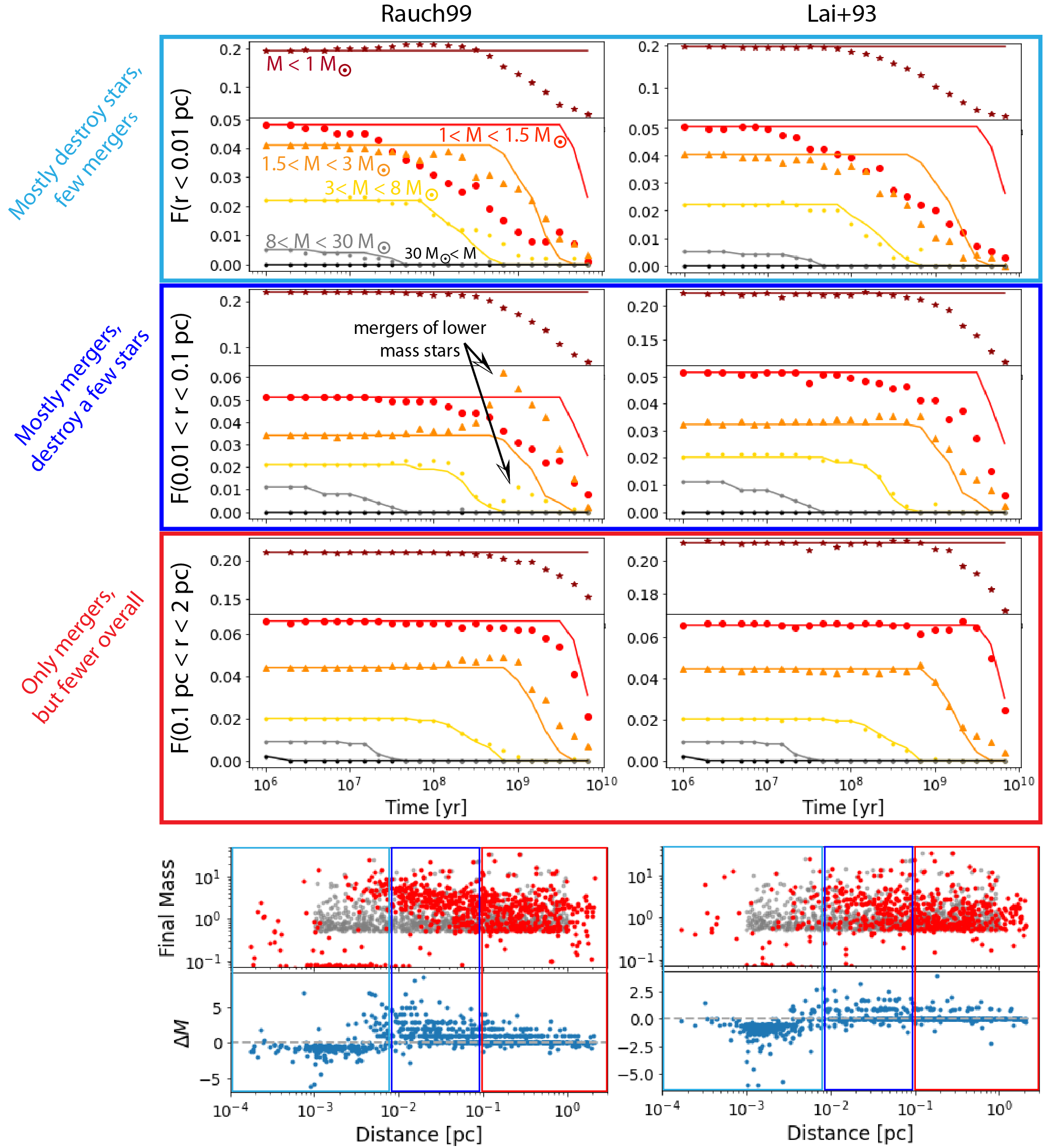}
    \caption{This figure juxtaposes the results for two different simulations, one that uses the equations from Rauch99 and the other, Lai+93. These simulations include relaxation, and the initial stellar masses are drawn from a Kroupa IMF. We assume a Bahcall-Wolf density profile for the surrounding stars. The first row shows the fraction of stars in each mass range as a function of time for the population within $0.01$ pc of the SMBH. The solid lines represent the expectation in the absence of collisions and relaxation. In other words, the solid lines decrease when stars within each mass range reach the end of their main-sequence lifetimes. The symbols show the simulated results. For example, as can be seen from the positions of the symbols relative to the curves, collisions destroy stars between $1$ and $3$~M$_\odot$ before they can evolve off the main-sequence. The second and third rows in the figure do the same for $0.01$-$0.1$~pc and $0.1$-$2$~pc from the SMBH. The last row shows the final masses and semimajor axes of the stars in red compared to the initial conditions, in grey, and the change in mass for the stars in blue. For a shallower profile ($\alpha=1.25$), see Figure \ref{fig:Nvstime125} in Appendix \ref{sec:moreresults}. } \label{fig:nvstime}
    \end{center}
\end{figure*}

Figure~\ref{fig:nvstime} shows the results of two simulations, Rauch99 on the left and Lai+93 on the right, with the aforementioned initial conditions. In the bottom row, we plot the final masses and semimajor axes of the stars in red and the initial masses and semimajor axes in grey. The bottom row also shows the change in mass of each star versus its initial position. For effective visualization, we divide the nuclear star cluster into three regions that roughly correspond to different collision outcomes based on Figure \ref{fig:timescales}.
The regimes are: \begin{itemize}
    \item 
$r < 0.01$~pc (top row, light blue border), where the velocity dispersion is high and we expect repeated collisions lead to mostly destruction of stars. 
\item $0.01< r < 0.1$~pc (middle row, dark blue border), where the velocity dispersion and relatively short collision timescale allows mergers to take place. 
\item  $0.1 < r < 2$~pc (third row, red border). In this regime, while the velocity dispersion allows mergers to take place, the collision timescale is longer than the stellar lifetimes. The number of collisions is therefore low.  
\end{itemize} 
These three rows in Figure \ref{fig:nvstime} depict the the number of stars within a certain mass range as a function of time. The solid lines represent the expectation in the absence of collisions. The solid lines decrease when stars within each mass range reach the end of their main-sequence lifetimes. The symbols show stellar abundances when collisions are included in the simulation. In the bottom row of the figure, we also include plots in the style of Figure~\ref{fig:SMMrlx} with the different regimes highlighted to guide the eye.


In the first region (top row), the populations of all the low mass stars decline over time. Collisions destroy these stars before the can evolve off the main-sequence. However, multiple collisions are required in order to destroy the stars. As can be observed in the figure, it takes about one billion years to halve the population of low-mass stars in both the Rauch99 and Lai+93 simulations. This half-life is about an order of magnitude longer than the collision timescale. We attribute this longevity to the fact that most collisions are not head-on and therefore less destructive. Additionally, stars losing mass through collisions can temporarily increase the number of stars in a lower mass range. For example, as $1$-$1.5$~M$_\odot$ stars lose mass, they will temporarily increase stars in the $0.5$-$1$~M$_\odot$ mass range.

Between $0.01$ and $0.1$~pc, the number of stars in the ranges $M<1$~M$_\odot$ and $1$-$1.5$~M$_\odot$ still decrease. However, while some of these stars may undergo destructive collisions, it is apparent from the final masses, as shown in the last row of the figure, that their changes in number are due mostly to mergers. This can be seen most clearly in the Rauch99 simulation, where there is a visible increase in the number of $1.5$-$3$~M$_\odot$ stars such that they outnumber the $1$-$1.5$~M$_\odot$ stars. Additionally, increasing the mass of the stars even slightly causes the stars to evolve off the main-sequence more quickly. As a result, the red symbols, for example, decrease more quickly compared to the expectation based on the original population (solid line). The Lai+93 prescription leads to more moderate mass growth than the Rauch99 one. As a result, the faster decline of the $1$-$1.5$~M$_\odot$ stars than expected is likely due to accelerated evolution, not stars leaving this mass range entirely. Lastly, in the $0.1 < r < 2$~pc, similar effects from mergers are present, but to a far lesser degree.

\begin{figure*}
\begin{center}
	\includegraphics[width=0.95\textwidth]{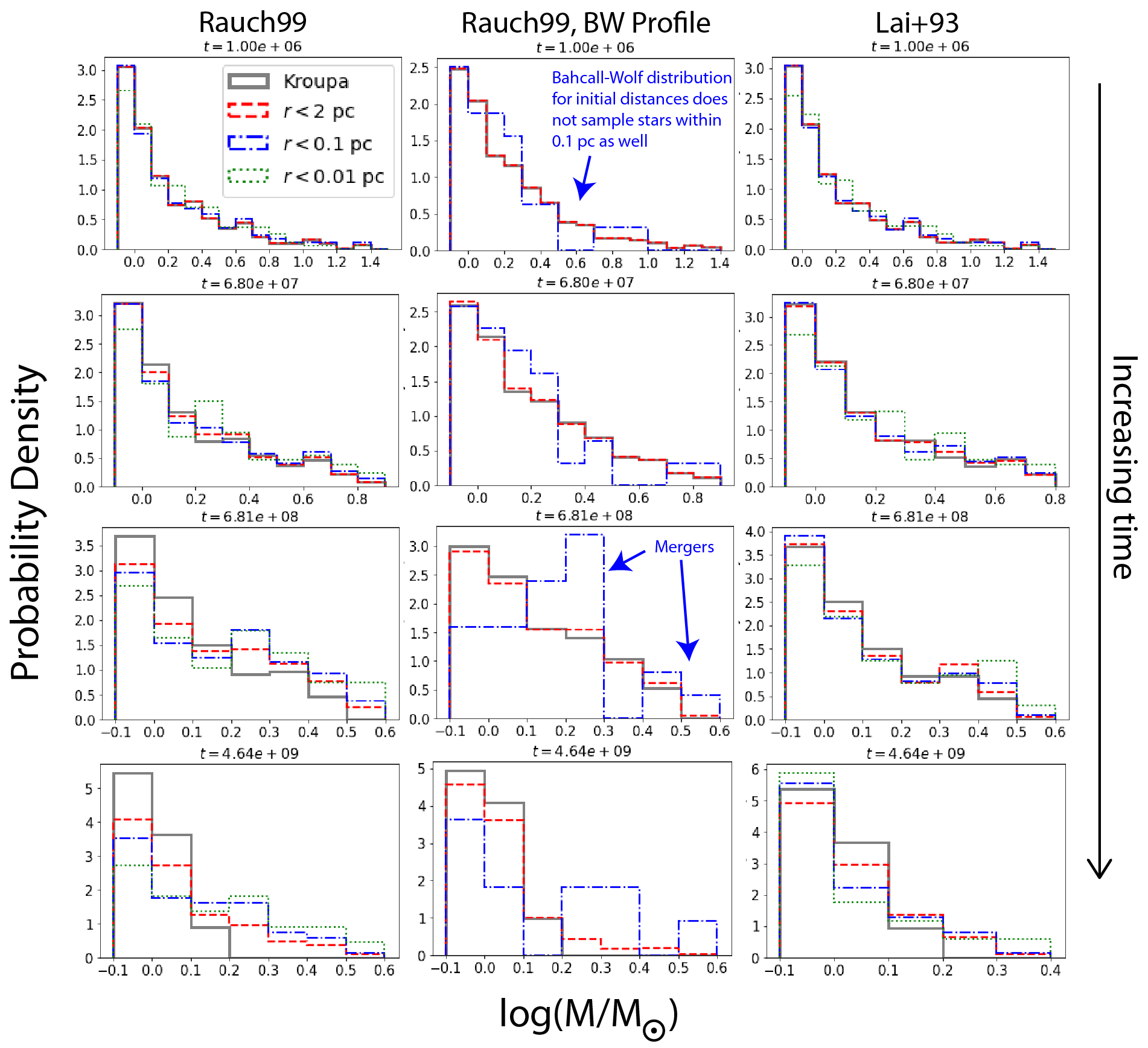}
    \caption{This figure shows the simulated stellar population's distribution of masses at a given time, noted in years at the top of each subplot. Each column corresponds to a different simulation, ordered chronologically from top to bottom. The right and left columns are from the same simulations shown in Figure~\ref{fig:nvstime}. The middle column shows the results of a simulation identical to the one on the left except the stars are sampled from a Bahcall-Wolf distribution; they are not distributed uniformly in log distance. The grey histograms show the expected surviving stars from a population that started with a Kroupa IMF. The red dashed, blue dash-dotted, and green dotted histograms correspond to the stars internal to $2$, $0.1$, and $0.01$~pc respectively.} \label{fig:mainhistogram}
    \end{center}
\end{figure*}

We suggest that collisions may result in more massive, young-seeming stars. In particular, because collisions need time to act, a massive stellar population may emerge after about a Gyr. This emergent population skews towards higher masses, yielding a mass function that resembles a top heavy one.  This behaviour is depicted in Figure \ref{fig:mainhistogram}, which shows stellar mass distribution for a few representative snapshots in time. In this Figure, we again show our two main simulations, also depicted in Figures~\ref{fig:SMMrlx} and \ref{fig:nvstime}, in the left and right columns. The simulation shown in the middle column is identical to the Rauch99 simulation on the left, except the initial distances from the SMBH of the $1000$ stars are drawn from a Bahcall-Wolf distribution ($\alpha = 1.75$). This initial condition does not adequately sample stars close to the SMBH. However, this simulation provides a check on mass functions from the other two simulations, which over represent stars close to the SMBH in population demographics because most of the stars in the nuclear star cluster are located further from the SMBH. Each subplot shows the mass function expected at a specific age for the population based on a Kroupa IMF in grey. The red dashed line shows the mass function at a given time for the entire cluster. The blue dash-dotted and green dashed lines show the mass distributions for the stars internal to $0.1$ and $0.01$ pc, respectively. We do not show the latter for the middle simulation because stars within this region are not adequately sampled for any meaningful result. While Figure~\ref{fig:SMMrlx} shows snapshots at only four different times, the expanded Figures~\ref{fig:histograms} and \ref{fig:gyrhistograms} in the Appendix display the full evolution of the mass function.

As shown in Figure~\ref{fig:SMMrlx}, the Rauch99 model, which favors mergers with minimal mass loss, leads to an excess of more massive stars in the inner $0.1$~pc (see bottom two rows of the left and middle columns). The Lai+93 column exhibits a similar effect, though to a lesser degree. These stars may appear rejuvenated and younger, especially compared to the surrounding cluster. Interestingly, the IMF of the young stars within $\sim 0.5$~pc of the SMBH has been suggested to be top heavy, with an excess of massive stars \citep[e.g.,][]{Paumard+06,Bartko+10,Lu+09,Lu+13}. Furthermore, the origins of young massive stars in the nuclear star cluster remain an open question; it remains uncertain whether they could have formed in situ in the extreme gravitational potential of the SMBH \citep[e.g.,][see \citet{Genzel+10rev} section 6 and \citet{Alexander05} section 7 for a review]{Morris93,Jackson+93,Gerhard01,Ghez+03,Genzel+03,Levin+03,Christopher+05,DaviesKing05,Paumard+06,Montero-Castano+09}. It has been suggested that stellar collisions and mergers may instead create these massive stars, explaining the so-called ``paradox of youth'' of the S-star cluster \citep[e.g.,][]{Ghez+03,Genzel+03,Antonini+10,Stephan+16,Stephan+19}. Our findings are consistent with a collisional formation scenario.

\section{Dust-Enshrouded G Objects} \label{sec:Gobjects}

Given the abundance of collisions within $0.1$~pc, here we consider what the merger products might look like post-collision, before the stellar object, a newly merged star, has had time to recollapse over a Kelven-Helmholtz timescale. In particular, these merged stars may have connections to gas and dust-enshrouded stellar objects, or G objects, observed in the GC  \citep[see][]{Ciurlo+20}. Several works in the literature have addressed the mysterious origin of the G object population \citep[e.g.,][]{Murray-Clay+12,Schartmann+12,Burkert+12,Zajacek+17,Madigan+17,Owen+23}. For example, stellar mergers from binary systems may explain the presence of these distended stellar objects \citep[e.g.,][]{Witzel+14,Prodan+15,Stephan+16,Stephan19}. In the nuclear star cluster, gravitational perturbations from the SMBH on a stellar binary's orbit can drive the binary to merge \citep[e.g.,][]{Antonini&Perets12,Prodan+15,Stephan+16,Stephan+19}. Collisions between other species like compact objects and RGs have also been invoked to explain the formation of G objects \citep{Mastrobuono-Battisti+21}.

\begin{figure}
	\includegraphics[width=0.99\columnwidth]{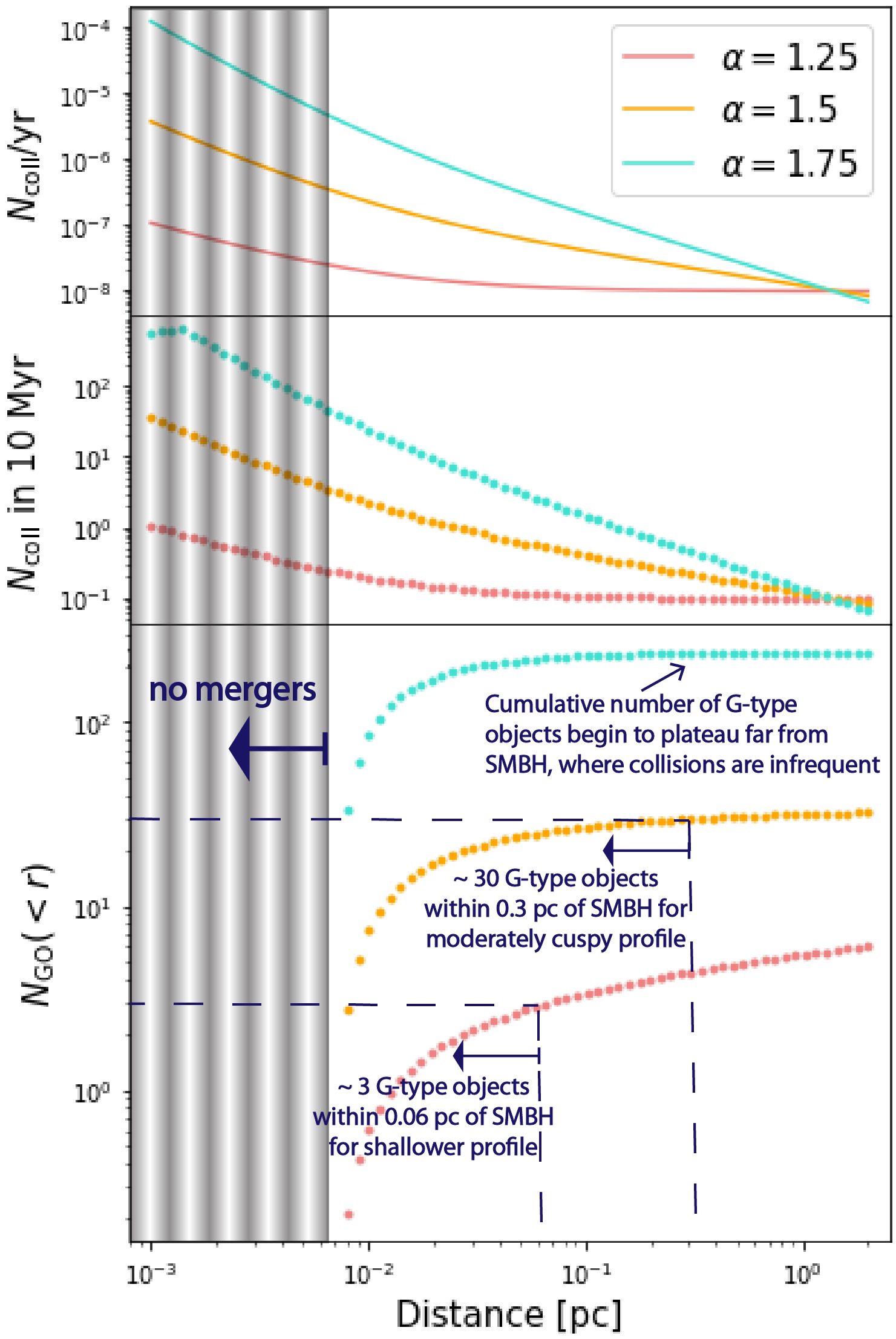}
	\caption{Here we provide predictions for the number of G-type objects in the nuclear star cluster formed through main-sequence stellar collisions. The uppermost panel shows the total number of stellar collisions per year as a function of distance from the SMBH for different density profiles. The second row shows the total number of collisions at each distance that occurs within 10 Myr, the approximate thermal or Kelvin-Helmholtz timescale of a Sun-like star. The last row shows the cumulative number of G-type objects within a given distance from the SMBH. An annotation is provided on the Figure to demonstrate how to interpret these curves.}
     \label{fig:GO}
\end{figure}

We estimate the number of G objects from main-sequence stellar collisions in the GC in Figure \ref{fig:GO}. We begin by calculating the number of collisions that occur per year at each distance from the SMBH. A post-collision merger product remains puffy and extended for a thermal timescale \citep[see][]{Stephan+16,Stephan+19}.\footnote{The time it takes for two merging stars to appear as a single, non-distended object is uncertain and may be shorter than the thermal timescale, $10^7$~yr \citep[e.g., COSMIC package from][]{Breivik+20,Ivanova11,Ivanova+13}. However, a more detailed analysis is beyond the scope of this paper.} Therefore, the number of G-type objects visible at a given distance from the SMBH is equal to the collision rate times $10^7$~yr, plotted in the second row of the Figure. The last row shows the cumulative number of G objects as a function of distance from the SMBH for three different stellar density profiles, $\alpha = 1.25$, $1.5$, and $1.75$. This analytic calculation assumes a uniform population of solar mass stars. We do not include collisions internal to about $0.008$~pc, based on Figures~\ref{fig:SMMrlx} and \ref{fig:nvstime}, because collisions in this region are largely destructive and do not result in mergers. For the least cuspy of the profiles considered, $\alpha = 1.25$, we expect approximately $10$ G objects, a number consistent with the current G object populations \citep{Ciurlo+20}. For the other two steeper density profiles, $\alpha = 1.5$ and $1.75$, G objects should number in the tens and hundreds, respectively. The density profile in the Galactic Center in present day appears to be shallower than a Bahcall-Wolf profile, lying between about $\alpha = 1.25$ and $1.5$ \citep[e.g.,][]{Genzel+03,Gallego-Cano+18}.

\section{Implications for the Red Giant Population} \label{sec:RGs}

Observations of the nuclear star cluster indicate that the surface number density of RGs plateaus within about $0.3$~pc of the SMBH \citep[e.g.,][]{Genzel+96,Buchholz+09,Do+09,Gallego-Cano+18,Habibi+19RG}. This plateau diverges from the cusp-like density profile expected for an older, evolved stellar population \citep[e.g.,][see the latter for the distribution expected from steady-state scatterings]{BahcallWolf76,LinialSari22}. The observed missing RG cusp has motivated several proposed mechanisms to destroy RGs near the SMBH, though none has been able to completely explain it \citep{Davies+98,Alexander99,BaileyDavies99,Dale+09,Zajacek+20}. However, it remains possible that collisions between main-sequence stars contribute to the dearth of RGs. As shown in Figure~\ref{fig:timescales}, $1$-$3$~M$_\odot$ stars within $\sim0.1$~pc of the SMBH are likely to experience a collision before evolving off the main-sequence.
To ensure a deficit in RGs, these stars must lose significant mass -- or be completely destroyed -- before they can evolve off the main-sequence, or mergers may accelerate their evolution off the main-sequence. 

Recently, \citet{Mastrobuono-Battisti+21} undertook a comprehensive N-body study of collisions between many different species in the GC and comment on a variety of applications, including the missing RGs. Their results suggest that collisions between main-sequence stars cannot explain the underabundance of RGs in the GC. They assume that the stars merge with no mass loss when the velocity dispersion is between $100$ and $300$~$\mathrm{km/s}$, as is the case in their model. However, a velocity dispersion of $300$~$\mathrm{km/s}$ corresponds to a distance of about $0.07$~pc from the SMBH. Furthermore, a $1$~M$_\odot$ star may diffuse into regions of larger velocity dispersion over its main-sequence lifetime, which is longer than the $\sim 1$~Gyr relaxation timescale. In this section, motivated by Figures~\ref{fig:nvstime} and \ref{fig:mainhistogram}, we assess whether collisions closer to the SMBH can affect the RG count in the GC.

\begin{figure}
	\includegraphics[width=0.99\columnwidth]{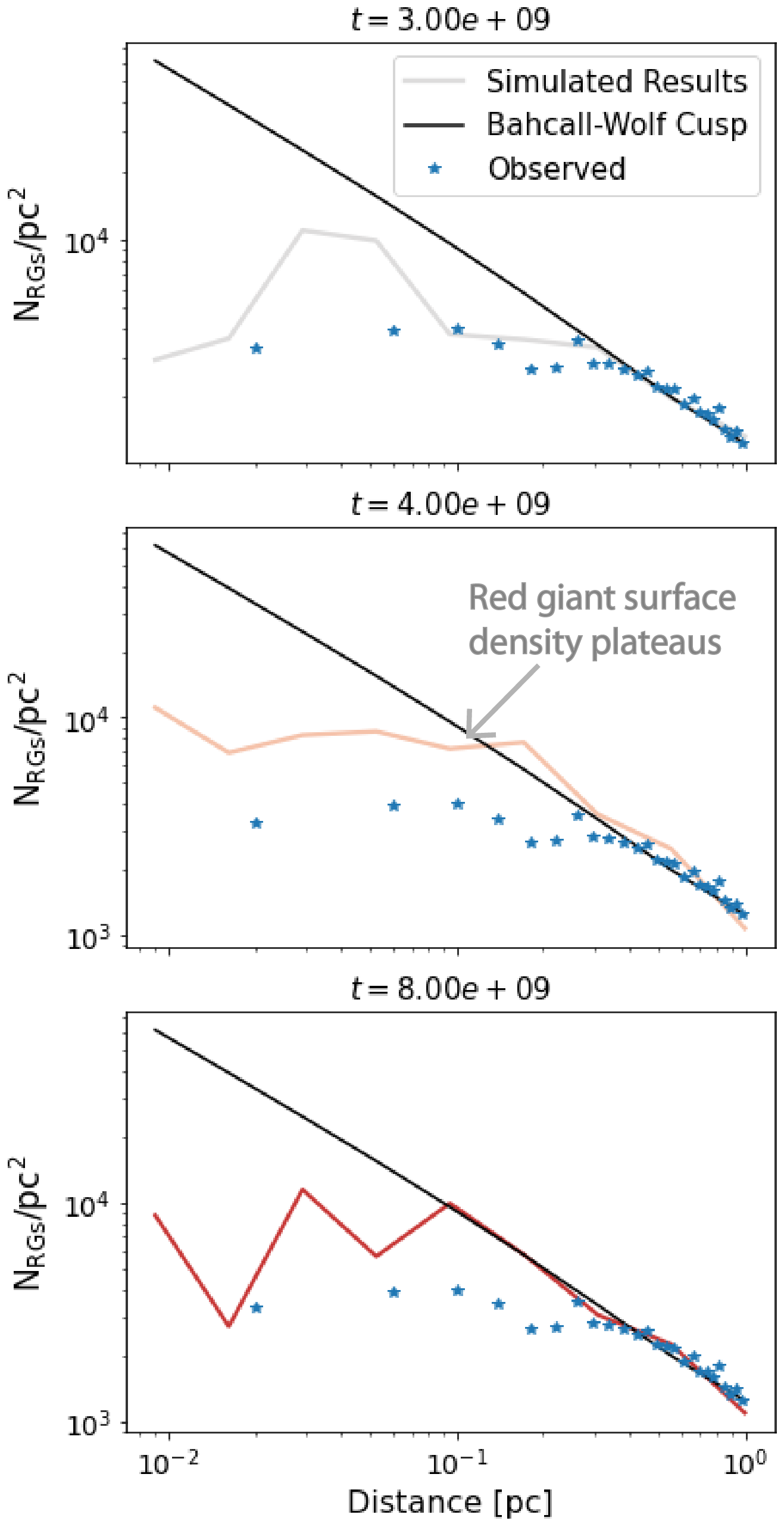}
	\caption{We generate surface density profiles for the RGs from our fiducial Rauch99 simulation, depicted in the left column of Figure~\ref{fig:nvstime}, for example. This figure shows examples of the simulated RG surface density at three different times: $3$~Gyr in grey in the top panel, $4$~Gyr in peach in the second panel, and $8$~Gyr in red in the bottom panel. The black line depicts a Bahcall-Wolf profile, the theoretical prediction for an old population \citep{BahcallWolf76}. The blue stars show the observed RG surface density from \citet{Gallego-Cano+18} figure 9. We scale our simulated profiles to roughly match the observed number of RGs at $1$~pc. The simulated RG surface densities, similar to the data, plateau within $\sim 1$~pc from the SMBH, diverging from the predicted cusp.}
     \label{fig:RGs}
\end{figure}

We generate RG surface density profiles at different snapshots in time from our fiducial Rauch99 simulation, shown for example in the left column of Figure~\ref{fig:nvstime}. For adequate resolution, we increase the number of stars in our sample from $1000$ to $4000$. Only a subset of this population will be RGs at any given time. The RGs used to generate the surface density profiles shown in Figure~\ref{fig:RGs} number at about $100$ each. We bin the number of RGs by distance from the SMBH. We then take the number of RGs formed per star and scale it by the number of stars expected at that distance, normalized using the M-$\sigma$ relation \citep{Tremaine+02}.
Figure~\ref{fig:RGs} shows the results at three different ages for the stellar population, $3$, $4$, and $8$~Gyr in grey, peach, and red, respectively. We compare these results to data from figure 9 of \citet{Gallego-Cano+18}, represented by the blue star-shaped dots. Extinction may explain the dip in the RG count data at $0.2$~pc \citep{Buchholz+09,Gallego-Cano+18}. We have scaled our simulated results to coincide roughly with the number of observed RGs at $1$~pc. In each panel, we also include a Bahcall-Wolf profile to guide the eye.

Both our simulated results and the observed profile  lack a cusp internal to about $0.1$~pc. In this region, mergers occur. As a result, these stars evolve off the main-sequence more rapidly. This process can boost the number of RGs at earlier times. The timing at which these stars evolve off the main-sequence has a radial dependence, based on the collision timescale. A population over a few Gyr old may have already experienced this accelerated boost of RGs, leading to a deficit in present time. Additionally, in this region, where multiple collisions are possible, many stars may undergo enough mergers to exceed $3$~M$_\odot$ (see also the last panel of Figure~\ref{fig:SMMrlx}). Lastly, more massive stars spend less time as RGs, so merging $1$~M$_\odot$ stars into $2$~M$_\odot$ stars, for example, reduces the chances of catching and observing them as RGs.

The discussion above pertains to stars outside $\sim 0.01$~pc, the innermost distance from the SMBH included in Figure~\ref{fig:RGs}. However, as seen in Figure~\ref{fig:nvstime}, stellar collisions destroy most low-mass ($\lesssim 3$~M$_\odot$) stars within $\sim 0.01$~pc of the SMBH before they can evolve off the main-sequence. Thus, we also expect a clear RG deficit in this region. We note that it remains uncertain whether stellar collisions, even at extremely high velocities, can destroy stellar cores. The first one or few collisions in this $\lesssim 0.01$~pc region can easily remove that outer layers of a star. In our simulations, later collisions can go on to destroy the low-mass collision remnants. However, if these stellar cores prove more resilient against collisions, our results may need to be revisited.

We stress that the simulated curves in Figure~\ref{fig:RGs} should not be used as predictions for the number of RGs at a given distance from the SMBH. This exercise is meant to draw attention to the general shape of the RG surface density profile based on our results. The actual RG density is likely a composite of several of these theoretical curves. The GC contains populations of different ages and its star formation history remains an open area of research \citep[e.g.,][]{Pfuhl+11,Schodel+20,Nogueras-Lara+21,Chen+23}. Additionally, our simulations consider an evolving sample of stars embedded in a fixed, unchanging cluster. In an actual stellar cluster, a merger between two stars results in a single object. Each collision should therefore reduce the number of particles in the system. In our simulations, however, the number of particles, or stars, remains the constant in our sample. Our final numbers may therefore over count by a factor of two. Lastly, in addition to the caveats mentioned above, we also assume an initial Bahcall-Wolf profile ($\alpha = 1.75$) for the stars. Two-body scattering in the presence of a black hole population in the nuclear star cluster may instead cause the stars to settle on a slightly shallower ($\alpha = 1.5$) profile within a critical radius from the SMBH \citep{LinialSari22}. A shallower initial profile may lead to greater deficit of RGs within $0.1$~pc. We reserve a more thorough examination of the surface density profile for future work. However, these results suggest that main-sequence stellar collisions merit further consideration in the context of the evolved population.

\section{Summary} \label{sec:discussion}
Most stars in the GC will experience direct collisions with other stars. In this study, we characterize the outcomes of these collisions as a function of distance from the SMBH. In particular, we consider three regimes: (1) within $\sim 0.01$~pc of the SMBH, where the velocity dispersion is large and collisions are destructive, (2) $0.01$-$0.1$~pc, where collisions are common and result in stellar mergers, and (3) outside of $\sim 0.1$~pc, where collisions do not occur very frequently, but always result in mergers. The details of each collision, specifically the amount of mass ejected during the collision, are more difficult to determine. In particular, the mass lost from the stars depends on their relative velocity, mass ratio, structure, and impact parameter \citep[e.g.,][]{Lai+93,FreitagBenz02}. 
We adopt different prescriptions to treat the mass loss, including a few based on fitting formulae from SPH studies \citep{Lai+93,Rauch99}. Our models also account for two-body relaxation, which works to change the semimajor axes and eccentricities of the stellar orbits in the GC over time. We find the following:

(i) \textit{The Formation of Massive Stars:} The mass loss prescriptions all lead to qualitatively similar results, consistent with our expectations for the different regimes. In particular, in the $0.01$-$0.1$~pc regime, stars with mass greater than $\sim 8$~M$_\odot$ can form from mergers between $1$~M$_\odot$ stars. Furthermore, in the case that collisions result in only moderate mass loss even at high impact velocities, as is the case for the Rauch99 equations, repeated collisions between low-mass can consistently form massive ($\gtrsim 10$~M$_\odot$) stars. These massive products are mainly localized within the $\sim 0.01$-$0.1$~pc region, as there is insufficient time for them to diffuse to distances $\sim 1$~pc from the SMBH. However, as can be seen in Figures~\ref{fig:SMMrlx} and \ref{fig:nvstime}, significant mass growth through mergers remains possible, albeit less probable, at $\lesssim 1$~pc from the SMBH. We predict that there should be at least $100$ massive stars formed from lower mass stars in this region. Generally, mergers may cause the mass function of the population to appear top-heavy (see Figure~\ref{fig:mainhistogram}, in keeping with observations of this region \citep[e.g.,][]{Paumard+06,Bartko+10,Lu+09,Lu+13}.

(ii) \textit{A Diffuse Population of Low Mass Remnants:} Collisions can divest many main-sequence stars of their outer layers, creating a low-mass stellar object. This result is consistent with previous studies \citep{Rauch99}. These collision remnants can migrate through relaxation processes from their place of formation, $\lesssim 0.01$~pc, to greater distances from the SMBH. We therefore predict a population of low-mass peculiar stars throughout the GC. Additionally, the mass ejected through grazing or off-axis collisions may produce gas and dust features in the Galactic Center, like the observed X7 \citep[e.g.,][]{Ciurlo+23}, whose distance from the SMBH is comparable to where $\sigma$ becomes $\lesssim v_{esc}$ (see Figure~\ref{fig:timescales}).

(iii) \textit{The Survival of Low-Mass Stars within $0.01$ pc:} While collisions within $0.01$~pc tend to be very destructive for the stars, our results suggest that their populations may persist for longer than expected based on the collision timescale alone, similar to findings in other studies \citep[e.g.,][]{Rauch99}. In fact, it takes $\sim 1$~Gyr to halve the populations of low-mass ($\lesssim 3$~M$_\odot$) stars. This timescale is the same order of magnitude as the relaxation timescale in this region (Figure~\ref{fig:timescales}). Episodes of star formation, not included in this study, may therefore help replenish the stellar population in this region. Depending on the age of the population, there may be an undetected population of low-mass stars and collision remnants in the innermost region of the nuclear star cluster.

(iv) \textit{G Objects:} Collisions between main-sequence stars may have connections to the G objects, stellar objects enshrouded in gas and dust observed in the GC \citep[e.g.,][]{Ciurlo+20}. A merger between two collided stars may produce a puffy G-like object that recollapses over the thermal timescale of a solar mass star, $10^7$~yr \citep[similarly, for binary mergers, see e.g.,][]{Stephan+19}. We provide predictions for the number of G type objects expected in the GC from stellar collisions given different stellar density profiles. For a stellar density profile with index $\alpha$ between $1.25$ and $1.5$, there may be $\sim 10$ G objects observable in the inner parsec of the GC.

(v) \textit{Connections to the Missing Red Giants:} Observations of the GC indicate a deficit of RGs within about $\sim 0.3$~pc of the SMBH \citep[e.g.,][]{Genzel+96,Buchholz+09,Do+09,Gallego-Cano+18,Habibi+19RG}. We consider whether main-sequence stellar collisions may help explain this observational puzzle. We find that within $\sim 0.01$~pc of the SMBH, stellar collisions destroy most low-mass stars before they can evolve off the main-sequence  (see Figure~\ref{fig:nvstime}). Thus, we expect a lack of RGs in this region. Further from the SMBH, between $\sim 0.01$-$0.1$~pc, mergers can result in fewer RGs. As seen in Figure~\ref{fig:RGs}, the RG surface density profile plateaus in this region, an effect similar to what is observed in the GC. Both the profiles from our models and observations diverge from the density cusp expected for an old, dynamically relaxed population: $\alpha = 1.75$ in Eq.~\ref{eq:density}.


\vspace{1em}

Collisions play a crucial role in shaping the stellar populations in the inner parsec of the GC. We show that this dynamical channel leads to a number of interesting observables, including rejuvenated, massive stars near the SMBH, G objects, and missing evolved stars. Our simulations focus on the dynamics of this dense environment using a statistical approach. We leverage previous SPH studies \citep[e.g.,][]{Lai+93} and intuitive toy models. However, our results highlight the need for future work to explore the rich interplay between hydrodynamics, stellar evolution, and collisions in these dense, dynamic environments.



\begin{acknowledgments}
We thank Anna Ciurlo, Cheyanne Shariat, Andrea Ghez, Jessica Lu, Mark Morris, Morgan MacLeod, and Enrico Ramirez-Ruiz for helpful discussion. SR thanks the Charles E. Young Fellowship, the Dissertation Year Fellowship, and the Thacher Fellowship for support. SR, SN, and RS acknowledge the partial support of the NSF-AST 2206428 grant. SR and SN acknowledge the partial support from NASA ATP  80NSSC20K0505. SN thanks Howard and Astrid Preston for their generous support. 
\end{acknowledgments}

\appendix

\section{The Structure of Our Sun: Basis for the Toy Core + Envelope Model} \label{sec:FML}

\begin{figure}
\begin{center}
	\includegraphics[width=0.4\columnwidth]{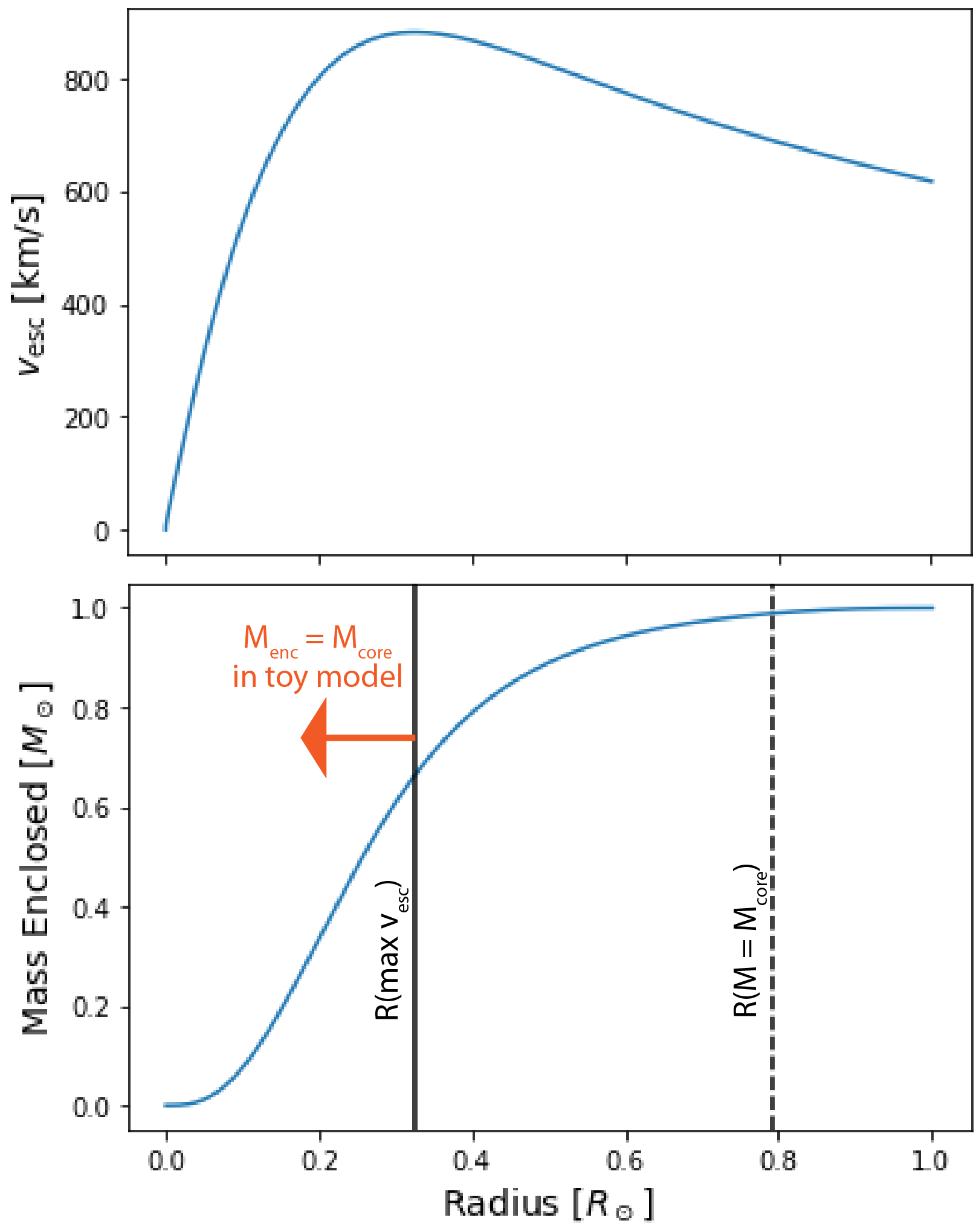}
	\caption{We base our toy ``core'' and ``envelope'' model on the structure of the Sun and generalize it to all stars. Above we show the escape speed as a function of radius within the star in the upper panel. It peaks around $0.33$~R$_\odot$ because most of the mass is concentrated at the center, as can be seen in the lower panel which plots mass enclosed versus radius. We define this radius as the extent of the ``core'' in our toy model, containing about $66 \%$ of the mass. We mark this radius with a solid black vertical line. Additionally, we show the radius of a $0.66$~M$_\odot$ main-sequence star, which will be the new radius of a solar mass star if it is completely divested of its ``envelope'' in a collision.}
     \label{fig:SSM}
\end{center}
\end{figure}

We base our toy ``core + envelope'' model on the structure of the Sun as described in \citet{C-Dalsgaard+96}.\footnote{https://users-phys.au.dk/$\sim$jcd/solar\_models/cptrho.l5bi.d.15c} The lower panel of Figure~\ref{fig:SSM} shows the mass enclosed within a given radius. Based on the mass enclosed within a given radius $r$, we calculate the escape velocity from a sphere with equivalent mass and radius, $v_\mathrm{esc} = \sqrt{2GM_\mathrm{enc}/r}$. Note that this quantity is not the true escape velocity from some point $r$ within the Sun as that would require integrating from $r$ to infinity to account for the outer layers.  We plot this quantity as a function of distance from the center of the Sun all the way to the surface in the upper panel of Figure~\ref{fig:SSM}. Because most of the Sun's mass is concentrated at its center, this quantity peaks at $0.33$~R$_\odot$, a radius which encloses $66 \%$ of the Sun's mass. We define the ``core'' of the star such that its radius coincides with this point. The ``envelope'' comprises the material external to this point, which is more loosely bound. Assuming that all stars have a similar structure, we scale this model to all star's within our considered mass range; we take $66 \%$ of the star's mass to be within $0.33 \times R_\mathrm{star}$ of the star's center and define this region as the ``core''. We then calculate collision outcomes as detailed in Section~\ref{sec:toycore}. This approach gives a straightforward, intuitive way of calculating the fractional mass loss and determining the outcome of a collision.



\section{Expanded Stellar Mass Function History Plots}

\begin{figure*}
\begin{center}
	\includegraphics[width=0.65\textwidth]{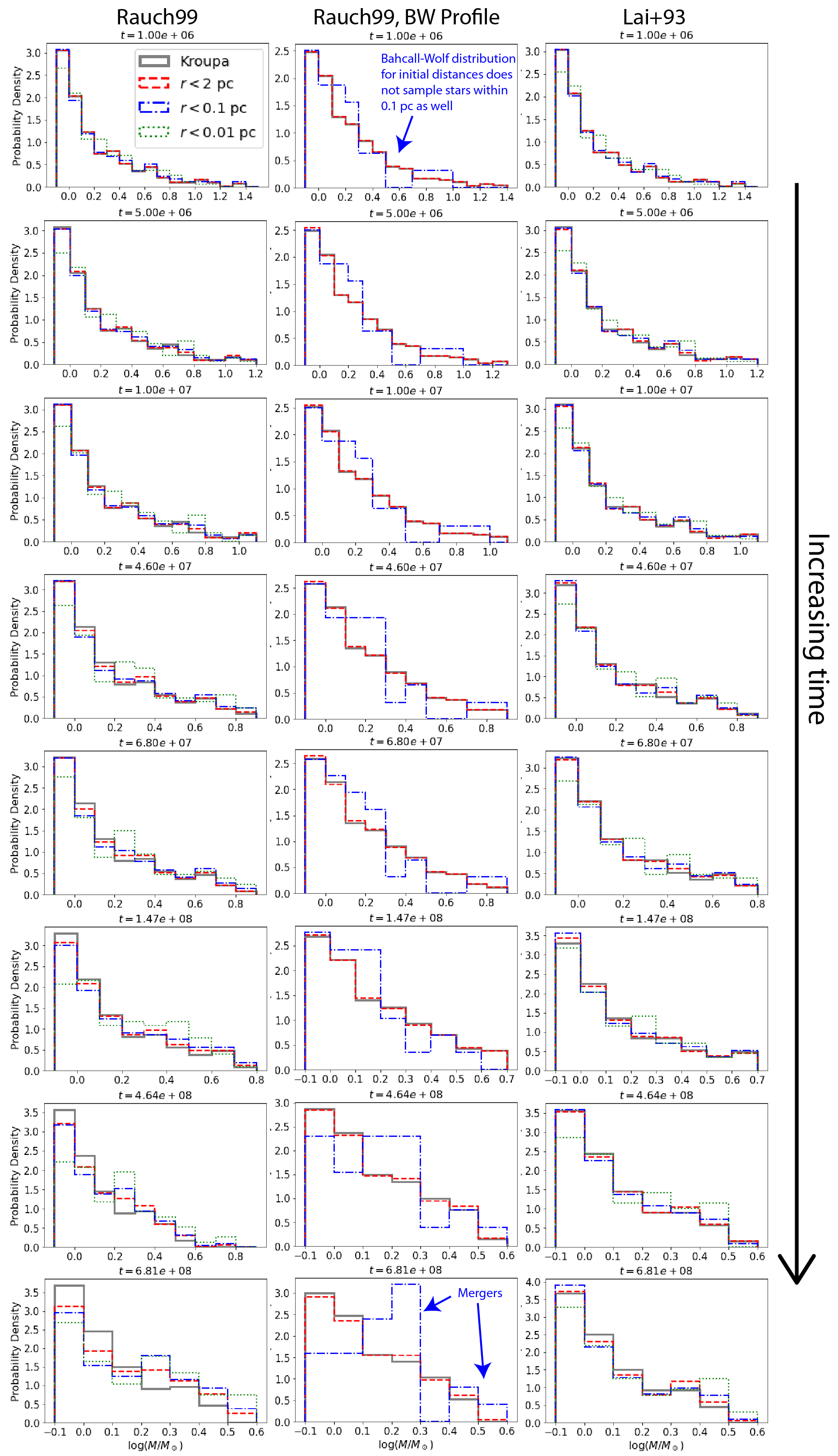}
    \caption{This Figure shows an expanded version of Figure~\ref{fig:mainhistogram} up to a Gyr. Each row corresponds to a snapshot in time, noted in years at the top of each plot. The left and right columns correspond to our Rauch99 and Lai+93 simulations, respectively. The middle panel shows results from a Rauch99 simulation that samples stars from a Bahcall-Wolf distribution ($\alpha = 1.75$).} \label{fig:histograms}
\end{center}
\end{figure*}

\begin{figure*}
\begin{center}
	\includegraphics[width=0.8\textwidth]{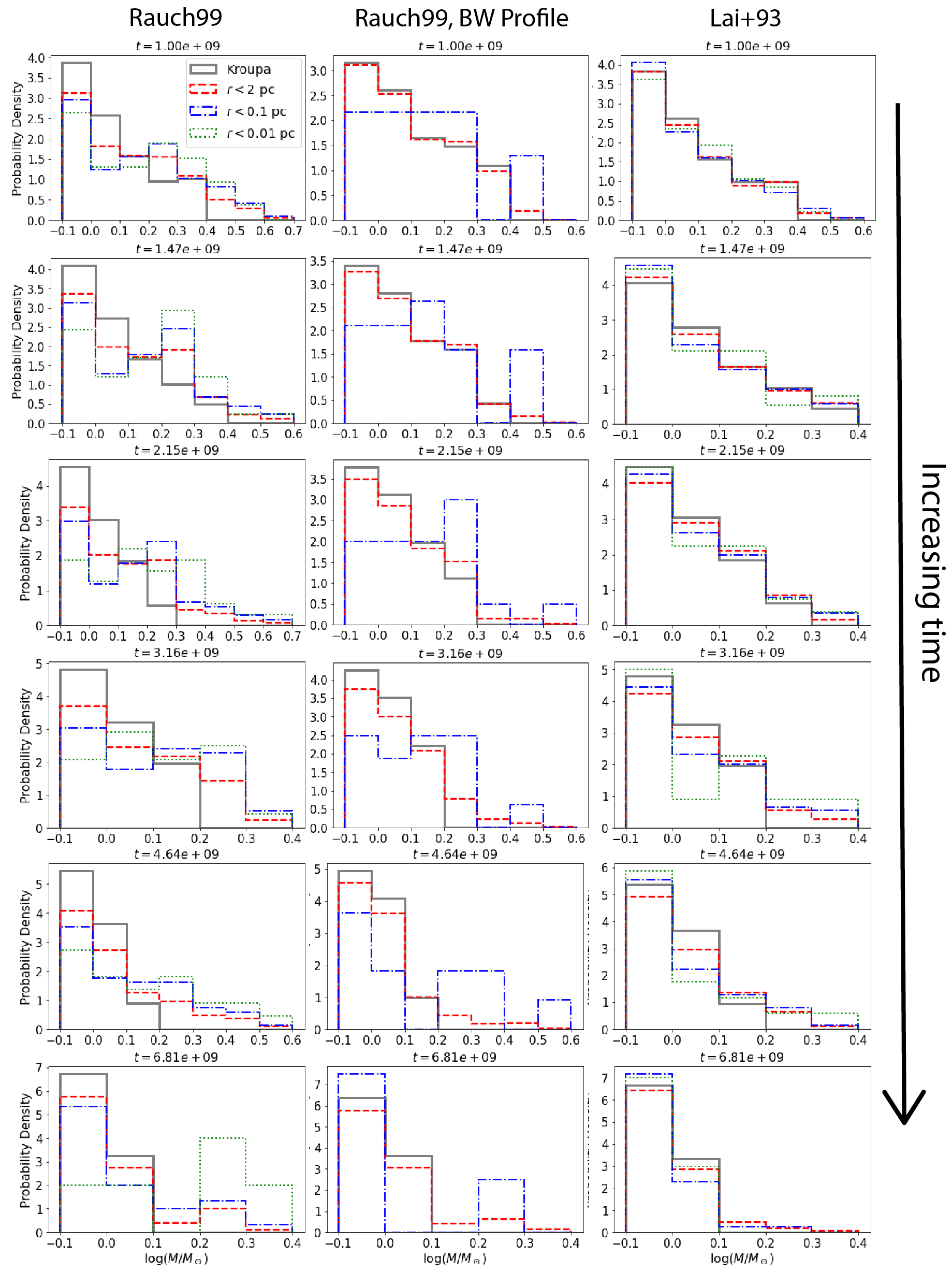}
    \caption{Showing stellar mass distributions from a Gyr onward, this figure is a continuation of Figure~\ref{fig:histograms}.} \label{fig:gyrhistograms}
 \end{center}   
\end{figure*}

In this section of the Appendix, we show expanded versions of Figure~\ref{fig:mainhistogram}, including more time steps to detail the evolution. Figure~\ref{fig:histograms} shows stellar mass distributions before a Gyr. Figure~\ref{fig:gyrhistograms} shows snapshots from a Gyr onward. Together, they provide a more detailed evolution of the mass function. Mergers skew the mass function toward higher masses, giving it a slightly top-heavy appearance.

\section{Other Stellar Density Profiles} \label{sec:moreresults}

Our simulations assume a background population of stars with some density profile that remains fixed. However, ovetime, we expect stellar collisions to modify the distribution, in particular by  eroding the cusp near the SMBH. In the absence of star formation, the new, shallower density profile can only lead to a lower collision rate. As a result, our predictions of the rate of destruction of stars may be overpredicting their decline in population. Additionally, two-body scattering with a population of heavier black holes may cause the stars to lie on a profile with index $\alpha = 1.25$ \citep{LinialSari22}, and observations of the GC's nuclear star cluster suggest that the profile may be shallower \citep[e.g.,][]{Genzel+03,Gallego-Cano+18}. Therefore, we are including a version of the Rauch99 simulation with an $\alpha = 1.25$ density profile for the surrounding star cluster, as opposed to our default $\alpha = 1.75$.
The most realistic picture probably lies between the $\alpha = 1.75$ and $\alpha = 1.25$ cases.

\begin{figure*}
\begin{center}
	\includegraphics[width=0.5\textwidth]{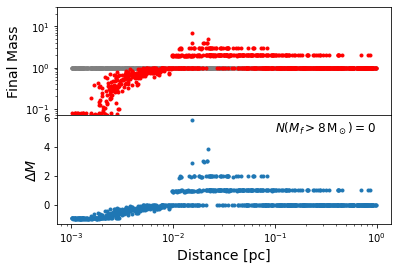}
    \caption{This figure shows the results of a Rauch99 simulation with a shallower $\alpha = 1.25$ density profile for the surrounding stars. Other than the density profile, this simulation has identical conditions to the one depicted in the last row of Figure~\ref{fig:SMM}: a uniform population of solar mass stars and no relaxation processes included. The initial conditions are shown in grey in the top panel, while the red points show the final masses of the stars and the blue points, their total change in mass over the course of the simulation.} \label{fig:SMM125}
 \end{center}   
\end{figure*}

We show results in the same style as Figures~\ref{fig:SMM} and \ref{fig:nvstime}. In Figure~\ref{fig:SMM125}, we plot the results of a Rauch99, $\alpha = 1.25$ that does not include the effects of relaxation. There are fewer mergers due to the shorter collision timescale. The merger products are therefore less massive compared to the $\alpha = 1.75$ case. Additionally, the stars are not destroyed as efficiently. As can be seen in the upper panel of Figure~\ref{fig:SMM125}, the low-mass collision remnants persist for longer closer to the SMBH.

\begin{figure*}
\begin{center}
	\includegraphics[width=0.5\textwidth]{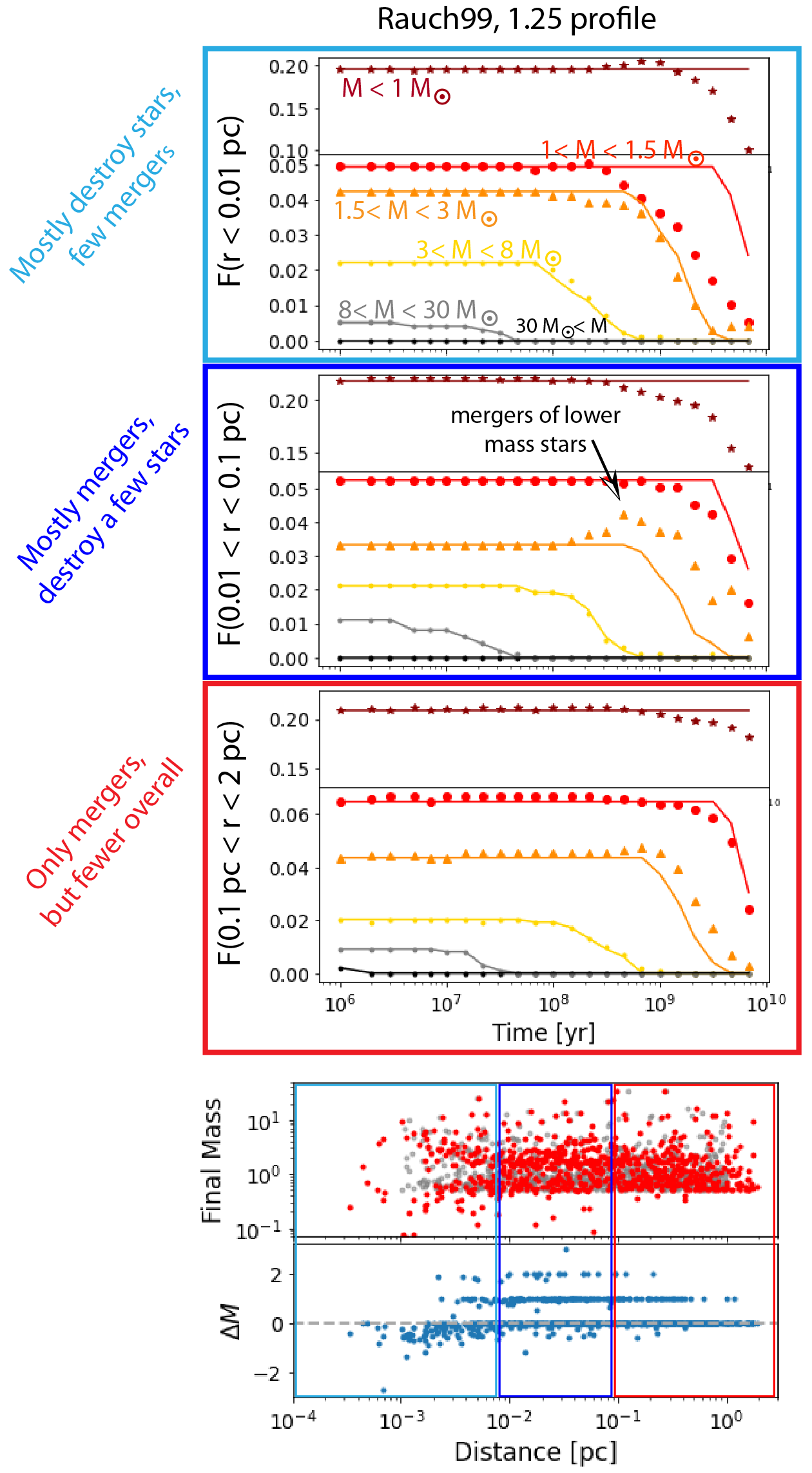}
    \caption{This figure has the same format as Figure~\ref{fig:nvstime}. The simulation used is identical to the left column of the aforementioned figure, except it uses a shallower $\alpha = 1.25$ density profile for the surrounding stars.} \label{fig:Nvstime125}
 \end{center}   
\end{figure*}

In a similar fashion, Figure~\ref{fig:Nvstime125} is identical to the left column of Figure~\ref{fig:nvstime} except for the assumed density profiles. More massive stars are still formed through collisions, but they are generally limited to less than $3$~M$_\odot$ because fewer collisions occur. For the same reason, the decline of the low-mass stars occurs more gradually than in the $\alpha = 1.75$ case.

\bibliography{sample631}{}

\begin{thebibliography}{}
\expandafter\ifx\csname natexlab\endcsname\relax\def\natexlab#1{#1}\fi
\providecommand{\url}[1]{\href{#1}{#1}}
\providecommand{\dodoi}[1]{doi:~\href{http://doi.org/#1}{\nolinkurl{#1}}}
\providecommand{\doeprint}[1]{\href{http://ascl.net/#1}{\nolinkurl{http://ascl.net/#1}}}
\providecommand{\doarXiv}[1]{\href{https://arxiv.org/abs/#1}{\nolinkurl{https://arxiv.org/abs/#1}}}

\bibitem[{{Aharon} \& {Perets}(2016)}]{AharonPerets16}
{Aharon}, D., \& {Perets}, H.~B. 2016, \apjl, 830, L1,
  \dodoi{10.3847/2041-8205/830/1/L1}

\bibitem[{{Alexander}(1999)}]{Alexander99}
{Alexander}, T. 1999, \apj, 527, 835, \dodoi{10.1086/308129}

\bibitem[{{Alexander}(2005)}]{Alexander05}
---. 2005, \physrep, 419, 65, \dodoi{10.1016/j.physrep.2005.08.002}

\bibitem[{{Alexander} \& {Hopman}(2009)}]{AlexanderHopman+09}
{Alexander}, T., \& {Hopman}, C. 2009, \apj, 697, 1861,
  \dodoi{10.1088/0004-637X/697/2/1861}

\bibitem[{{Alexander} \& {Pfuhl}(2014)}]{AlexanderPfuhl14}
{Alexander}, T., \& {Pfuhl}, O. 2014, \apj, 780, 148,
  \dodoi{10.1088/0004-637X/780/2/148}

\bibitem[{{Amaro Seoane}(2023)}]{AmaroSeoane23}
{Amaro Seoane}, P. 2023, arXiv e-prints, arXiv:2302.00014,
  \dodoi{10.48550/arXiv.2302.00014}

\bibitem[{{Antonini} {et~al.}(2010){Antonini}, {Faber}, {Gualandris}, \&
  {Merritt}}]{Antonini+10}
{Antonini}, F., {Faber}, J., {Gualandris}, A., \& {Merritt}, D. 2010, \apj,
  713, 90, \dodoi{10.1088/0004-637X/713/1/90}

\bibitem[{{Antonini} \& {Perets}(2012)}]{Antonini&Perets12}
{Antonini}, F., \& {Perets}, H.~B. 2012, \apj, 757, 27,
  \dodoi{10.1088/0004-637X/757/1/27}

\bibitem[{{Arca Sedda}(2020)}]{ArcaSedda20}
{Arca Sedda}, M. 2020, \apj, 891, 47, \dodoi{10.3847/1538-4357/ab723b}

\bibitem[{{Bahcall} \& {Wolf}(1976)}]{BahcallWolf76}
{Bahcall}, J.~N., \& {Wolf}, R.~A. 1976, \apj, 209, 214, \dodoi{10.1086/154711}

\bibitem[{{Bailey} \& {Davies}(1999)}]{BaileyDavies99}
{Bailey}, V.~C., \& {Davies}, M.~B. 1999, \mnras, 308, 257,
  \dodoi{10.1046/j.1365-8711.1999.02740.x}

\bibitem[{{Balberg} {et~al.}(2013){Balberg}, {Sari}, \& {Loeb}}]{Balberg+13}
{Balberg}, S., {Sari}, R., \& {Loeb}, A. 2013, \mnras, 434, L26,
  \dodoi{10.1093/mnrasl/slt071}

\bibitem[{{Bar-Or} {et~al.}(2013){Bar-Or}, {Kupi}, \& {Alexander}}]{Bar-Or+13}
{Bar-Or}, B., {Kupi}, G., \& {Alexander}, T. 2013, \apj, 764, 52,
  \dodoi{10.1088/0004-637X/764/1/52}

\bibitem[{{Bartko} {et~al.}(2010){Bartko}, {Martins}, {Trippe}, {Fritz},
  {Genzel}, {Ott}, {Eisenhauer}, {Gillessen}, {Paumard}, {Alexander},
  {Dodds-Eden}, {Gerhard}, {Levin}, {Mascetti}, {Nayakshin}, {Perets},
  {Perrin}, {Pfuhl}, {Reid}, {Rouan}, {Zilka}, \& {Sternberg}}]{Bartko+10}
{Bartko}, H., {Martins}, F., {Trippe}, S., {et~al.} 2010, \apj, 708, 834,
  \dodoi{10.1088/0004-637X/708/1/834}

\bibitem[{{Belczynski} {et~al.}(2020){Belczynski}, {Hirschi}, {Kaiser}, {Liu},
  {Casares}, {Lu}, {O'Shaughnessy}, {Heger}, {Justham}, \&
  {Soria}}]{Belczynski20}
{Belczynski}, K., {Hirschi}, R., {Kaiser}, E.~A., {et~al.} 2020, \apj, 890,
  113, \dodoi{10.3847/1538-4357/ab6d77}

\bibitem[{{Benz} \& {Hills}(1987)}]{BenzHills87}
{Benz}, W., \& {Hills}, J.~G. 1987, \apj, 323, 614, \dodoi{10.1086/165857}

\bibitem[{{Binney} \& {Tremaine}(2008)}]{BinneyTremaine}
{Binney}, J., \& {Tremaine}, S. 2008, {Galactic Dynamics: Second Edition}

\bibitem[{{Bradnick} {et~al.}(2017){Bradnick}, {Mandel}, \&
  {Levin}}]{Bradnick+17}
{Bradnick}, B., {Mandel}, I., \& {Levin}, Y. 2017, \mnras, 469, 2042,
  \dodoi{10.1093/mnras/stx1007}

\bibitem[{{Breivik} {et~al.}(2020){Breivik}, {Coughlin}, {Zevin}, {Rodriguez},
  {Kremer}, {Ye}, {Andrews}, {Kurkowski}, {Digman}, {Larson}, \&
  {Rasio}}]{Breivik+20}
{Breivik}, K., {Coughlin}, S., {Zevin}, M., {et~al.} 2020, \apj, 898, 71,
  \dodoi{10.3847/1538-4357/ab9d85}

\bibitem[{{Buchholz} {et~al.}(2009){Buchholz}, {Sch{\"o}del}, \&
  {Eckart}}]{Buchholz+09}
{Buchholz}, R.~M., {Sch{\"o}del}, R., \& {Eckart}, A. 2009, \aap, 499, 483,
  \dodoi{10.1051/0004-6361/200811497}

\bibitem[{{Burkert} {et~al.}(2012){Burkert}, {Schartmann}, {Alig}, {Gillessen},
  {Genzel}, {Fritz}, \& {Eisenhauer}}]{Burkert+12}
{Burkert}, A., {Schartmann}, M., {Alig}, C., {et~al.} 2012, \apj, 750, 58,
  \dodoi{10.1088/0004-637X/750/1/58}

\bibitem[{{Chen} {et~al.}(2023){Chen}, {Do}, {Ghez}, {Hosek},
  {Feldmeier-Krause}, {Chu}, {Bentley}, {Lu}, \& {Morris}}]{Chen+23}
{Chen}, Z., {Do}, T., {Ghez}, A.~M., {et~al.} 2023, \apj, 944, 79,
  \dodoi{10.3847/1538-4357/aca8ad}

\bibitem[{{Christensen-Dalsgaard}
  {et~al.}(1996{\natexlab{a}}){Christensen-Dalsgaard}, {Dappen}, {Ajukov},
  {Anderson}, {Antia}, {Basu}, {Baturin}, {Berthomieu}, {Chaboyer}, {Chitre},
  {Cox}, {Demarque}, {Donatowicz}, {Dziembowski}, {Gabriel}, {Gough},
  {Guenther}, {Guzik}, {Harvey}, {Hill}, {Houdek}, {Iglesias}, {Kosovichev},
  {Leibacher}, {Morel}, {Proffitt}, {Provost}, {Reiter}, {Rhodes}, {Rogers},
  {Roxburgh}, {Thompson}, \& {Ulrich}}]{Christensen-Dalsgaard+96}
{Christensen-Dalsgaard}, J., {Dappen}, W., {Ajukov}, S.~V., {et~al.}
  1996{\natexlab{a}}, Science, 272, 1286, \dodoi{10.1126/science.272.5266.1286}

\bibitem[{{Christensen-Dalsgaard}
  {et~al.}(1996{\natexlab{b}}){Christensen-Dalsgaard}, {Dappen}, {Ajukov},
  {Anderson}, {Antia}, {Basu}, {Baturin}, {Berthomieu}, {Chaboyer}, {Chitre},
  {Cox}, {Demarque}, {Donatowicz}, {Dziembowski}, {Gabriel}, {Gough},
  {Guenther}, {Guzik}, {Harvey}, {Hill}, {Houdek}, {Iglesias}, {Kosovichev},
  {Leibacher}, {Morel}, {Proffitt}, {Provost}, {Reiter}, {Rhodes}, {Rogers},
  {Roxburgh}, {Thompson}, \& {Ulrich}}]{C-Dalsgaard+96}
---. 1996{\natexlab{b}}, Science, 272, 1286,
  \dodoi{10.1126/science.272.5266.1286}

\bibitem[{{Christopher} {et~al.}(2005){Christopher}, {Scoville}, {Stolovy}, \&
  {Yun}}]{Christopher+05}
{Christopher}, M.~H., {Scoville}, N.~Z., {Stolovy}, S.~R., \& {Yun}, M.~S.
  2005, \apj, 622, 346, \dodoi{10.1086/427911}

\bibitem[{{Ciurlo} {et~al.}(2020){Ciurlo}, {Campbell}, {Morris}, {Do}, {Ghez},
  {Hees}, {Sitarski}, {Kosmo O'Neil}, {Chu}, {Martinez}, {Naoz}, \&
  {Stephan}}]{Ciurlo+20}
{Ciurlo}, A., {Campbell}, R.~D., {Morris}, M.~R., {et~al.} 2020, \nat, 577,
  337, \dodoi{10.1038/s41586-019-1883-y}

\bibitem[{{Ciurlo} {et~al.}(2023){Ciurlo}, {Campbell}, {Morris}, {Do}, {Ghez},
  {Becklin}, {Bentley}, {Chu}, {Gautam}, {Gursahani}, {Hees}, {O'Neil}, {Lu},
  {Martinez}, {Naoz}, {Sakai}, \& {Sch{\"o}del}}]{Ciurlo+23}
---. 2023, \apj, 944, 136, \dodoi{10.3847/1538-4357/acb344}

\bibitem[{{Dale} \& {Davies}(2006)}]{DaleDavies}
{Dale}, J.~E., \& {Davies}, M.~B. 2006, \mnras, 366, 1424,
  \dodoi{10.1111/j.1365-2966.2005.09937.x}

\bibitem[{{Dale} {et~al.}(2009){Dale}, {Davies}, {Church}, \&
  {Freitag}}]{Dale+09}
{Dale}, J.~E., {Davies}, M.~B., {Church}, R.~P., \& {Freitag}, M. 2009, \mnras,
  393, 1016, \dodoi{10.1111/j.1365-2966.2008.14254.x}

\bibitem[{{David} {et~al.}(1987{\natexlab{a}}){David}, {Durisen}, \&
  {Cohn}}]{David+87a}
{David}, L.~P., {Durisen}, R.~H., \& {Cohn}, H.~N. 1987{\natexlab{a}}, \apj,
  313, 556, \dodoi{10.1086/164997}

\bibitem[{{David} {et~al.}(1987{\natexlab{b}}){David}, {Durisen}, \&
  {Cohn}}]{David+87b}
---. 1987{\natexlab{b}}, \apj, 316, 505, \dodoi{10.1086/165222}

\bibitem[{{Davies} {et~al.}(1998){Davies}, {Blackwell}, {Bailey}, \&
  {Sigurdsson}}]{Davies+98}
{Davies}, M.~B., {Blackwell}, R., {Bailey}, V.~C., \& {Sigurdsson}, S. 1998,
  \mnras, 301, 745, \dodoi{10.1046/j.1365-8711.1998.02027.x}

\bibitem[{{Davies} \& {King}(2005)}]{DaviesKing05}
{Davies}, M.~B., \& {King}, A. 2005, \apjl, 624, L25, \dodoi{10.1086/430308}

\bibitem[{{Do} {et~al.}(2009){Do}, {Ghez}, {Morris}, {Lu}, {Matthews}, {Yelda},
  \& {Larkin}}]{Do+09}
{Do}, T., {Ghez}, A.~M., {Morris}, M.~R., {et~al.} 2009, \apj, 703, 1323,
  \dodoi{10.1088/0004-637X/703/2/1323}

\bibitem[{{Duncan} \& {Shapiro}(1983)}]{DuncanShapiro83}
{Duncan}, M.~J., \& {Shapiro}, S.~L. 1983, \apj, 268, 565,
  \dodoi{10.1086/160980}

\bibitem[{{Ferrarese} \& {Ford}(2005)}]{FerrareseFord05}
{Ferrarese}, L., \& {Ford}, H. 2005, \ssr, 116, 523,
  \dodoi{10.1007/s11214-005-3947-6}

\bibitem[{{Fregeau} \& {Rasio}(2007)}]{FregeauRasio07}
{Fregeau}, J.~M., \& {Rasio}, F.~A. 2007, \apj, 658, 1047,
  \dodoi{10.1086/511809}

\bibitem[{{Freitag} \& {Benz}(2002)}]{FreitagBenz02}
{Freitag}, M., \& {Benz}, W. 2002, \aap, 394, 345,
  \dodoi{10.1051/0004-6361:20021142}

\bibitem[{{Freitag} \& {Benz}(2005)}]{FreitagBenz}
---. 2005, \mnras, 358, 1133, \dodoi{10.1111/j.1365-2966.2005.08770.x}

\bibitem[{{Gallego-Cano} {et~al.}(2018){Gallego-Cano}, {Sch{\"o}del}, {Dong},
  {Nogueras-Lara}, {Gallego-Calvente}, {Amaro-Seoane}, \&
  {Baumgardt}}]{Gallego-Cano+18}
{Gallego-Cano}, E., {Sch{\"o}del}, R., {Dong}, H., {et~al.} 2018, \aap, 609,
  A26, \dodoi{10.1051/0004-6361/201730451}

\bibitem[{{Gallego-Cano} {et~al.}(2020){Gallego-Cano}, {Sch{\"o}del},
  {Nogueras-Lara}, {Dong}, {Shahzamanian}, {Fritz}, {Gallego-Calvente}, \&
  {Neumayer}}]{Gallego+20}
{Gallego-Cano}, E., {Sch{\"o}del}, R., {Nogueras-Lara}, F., {et~al.} 2020,
  \aap, 634, A71, \dodoi{10.1051/0004-6361/201935303}

\bibitem[{{Genzel} {et~al.}(2010{\natexlab{a}}){Genzel}, {Eisenhauer}, \&
  {Gillessen}}]{Genzel+10}
{Genzel}, R., {Eisenhauer}, F., \& {Gillessen}, S. 2010{\natexlab{a}}, Reviews
  of Modern Physics, 82, 3121, \dodoi{10.1103/RevModPhys.82.3121}

\bibitem[{{Genzel} {et~al.}(2010{\natexlab{b}}){Genzel}, {Eisenhauer}, \&
  {Gillessen}}]{Genzel+10rev}
---. 2010{\natexlab{b}}, Reviews of Modern Physics, 82, 3121,
  \dodoi{10.1103/RevModPhys.82.3121}

\bibitem[{{Genzel} {et~al.}(1996){Genzel}, {Thatte}, {Krabbe}, {Kroker}, \&
  {Tacconi-Garman}}]{Genzel+96}
{Genzel}, R., {Thatte}, N., {Krabbe}, A., {Kroker}, H., \& {Tacconi-Garman},
  L.~E. 1996, \apj, 472, 153, \dodoi{10.1086/178051}

\bibitem[{{Genzel} {et~al.}(2003){Genzel}, {Sch{\"o}del}, {Ott}, {Eisenhauer},
  {Hofmann}, {Lehnert}, {Eckart}, {Alexander}, {Sternberg}, {Lenzen},
  {Cl{\'e}net}, {Lacombe}, {Rouan}, {Renzini}, \& {Tacconi-Garman}}]{Genzel+03}
{Genzel}, R., {Sch{\"o}del}, R., {Ott}, T., {et~al.} 2003, \apj, 594, 812,
  \dodoi{10.1086/377127}

\bibitem[{{Gerhard}(2001)}]{Gerhard01}
{Gerhard}, O. 2001, \apjl, 546, L39, \dodoi{10.1086/318054}

\bibitem[{{Ghez} {et~al.}(2005){Ghez}, {Salim}, {Hornstein}, {Tanner}, {Lu},
  {Morris}, {Becklin}, \& {Duch{\^e}ne}}]{Ghez+05}
{Ghez}, A.~M., {Salim}, S., {Hornstein}, S.~D., {et~al.} 2005, \apj, 620, 744,
  \dodoi{10.1086/427175}

\bibitem[{{Ghez} {et~al.}(2003){Ghez}, {Duch{\^e}ne}, {Matthews}, {Hornstein},
  {Tanner}, {Larkin}, {Morris}, {Becklin}, {Salim}, {Kremenek}, {Thompson},
  {Soifer}, {Neugebauer}, \& {McLean}}]{Ghez+03}
{Ghez}, A.~M., {Duch{\^e}ne}, G., {Matthews}, K., {et~al.} 2003, \apjl, 586,
  L127, \dodoi{10.1086/374804}

\bibitem[{{Ghez} {et~al.}(2008){Ghez}, {Salim}, {Weinberg}, {Lu}, {Do}, {Dunn},
  {Matthews}, {Morris}, {Yelda}, {Becklin}, {Kremenek}, {Milosavljevic}, \&
  {Naiman}}]{Ghez+08}
{Ghez}, A.~M., {Salim}, S., {Weinberg}, N.~N., {et~al.} 2008, \apj, 689, 1044,
  \dodoi{10.1086/592738}

\bibitem[{{Gillessen} {et~al.}(2009){Gillessen}, {Eisenhauer}, {Trippe},
  {Alexand er}, {Genzel}, {Martins}, \& {Ott}}]{Gillessen+09}
{Gillessen}, S., {Eisenhauer}, F., {Trippe}, S., {et~al.} 2009, \apj, 692,
  1075, \dodoi{10.1088/0004-637X/692/2/1075}

\bibitem[{{Gillessen} {et~al.}(2017){Gillessen}, {Plewa}, {Eisenhauer}, {Sari},
  {Waisberg}, {Habibi}, {Pfuhl}, {George}, {Dexter}, {von Fellenberg}, {Ott},
  \& {Genzel}}]{Gillessen+17}
{Gillessen}, S., {Plewa}, P.~M., {Eisenhauer}, F., {et~al.} 2017, \apj, 837,
  30, \dodoi{10.3847/1538-4357/aa5c41}

\bibitem[{{Habibi} {et~al.}(2019){Habibi}, {Gillessen}, {Pfuhl}, {Eisenhauer},
  {Plewa}, {von Fellenberg}, {Widmann}, {Ott}, {Gao}, {Waisberg},
  {Baub{\"o}ck}, {Jimenez-Rosales}, {Dexter}, {de Zeeuw}, \&
  {Genzel}}]{Habibi+19RG}
{Habibi}, M., {Gillessen}, S., {Pfuhl}, O., {et~al.} 2019, \apjl, 872, L15,
  \dodoi{10.3847/2041-8213/ab03cf}

\bibitem[{{Heger} {et~al.}(2003){Heger}, {Fryer}, {Woosley}, {Langer}, \&
  {Hartmann}}]{Heger+03}
{Heger}, A., {Fryer}, C.~L., {Woosley}, S.~E., {Langer}, N., \& {Hartmann},
  D.~H. 2003, \apj, 591, 288, \dodoi{10.1086/375341}

\bibitem[{{Hoang} {et~al.}(2018){Hoang}, {Naoz}, {Kocsis}, {Rasio}, \&
  {Dosopoulou}}]{Hoang+18}
{Hoang}, B.-M., {Naoz}, S., {Kocsis}, B., {Rasio}, F.~A., \& {Dosopoulou}, F.
  2018, \apj, 856, 140, \dodoi{10.3847/1538-4357/aaafce}

\bibitem[{{Hoang} {et~al.}(2020){Hoang}, {Naoz}, \& {Kremer}}]{Hoang+20}
{Hoang}, B.-M., {Naoz}, S., \& {Kremer}, K. 2020, \apj, 903, 8,
  \dodoi{10.3847/1538-4357/abb66a}

\bibitem[{{Hurley} {et~al.}(2002){Hurley}, {Tout}, \& {Pols}}]{Hurley02}
{Hurley}, J.~R., {Tout}, C.~A., \& {Pols}, O.~R. 2002, \mnras, 329, 897,
  \dodoi{10.1046/j.1365-8711.2002.05038.x}

\bibitem[{{Ivanova}(2011)}]{Ivanova11}
{Ivanova}, N. 2011, \apj, 730, 76, \dodoi{10.1088/0004-637X/730/2/76}

\bibitem[{{Ivanova} {et~al.}(2013){Ivanova}, {Justham}, {Chen}, {De Marco},
  {Fryer}, {Gaburov}, {Ge}, {Glebbeek}, {Han}, {Li}, {Lu}, {Marsh},
  {Podsiadlowski}, {Potter}, {Soker}, {Taam}, {Tauris}, {van den Heuvel}, \&
  {Webbink}}]{Ivanova+13}
{Ivanova}, N., {Justham}, S., {Chen}, X., {et~al.} 2013, \aapr, 21, 59,
  \dodoi{10.1007/s00159-013-0059-2}

\bibitem[{{Jackson} {et~al.}(1993){Jackson}, {Geis}, {Genzel}, {Harris},
  {Madden}, {Poglitsch}, {Stacey}, \& {Townes}}]{Jackson+93}
{Jackson}, J.~M., {Geis}, N., {Genzel}, R., {et~al.} 1993, \apj, 402, 173,
  \dodoi{10.1086/172120}

\bibitem[{{Keshet} {et~al.}(2009){Keshet}, {Hopman}, \&
  {Alexander}}]{Keshet+09}
{Keshet}, U., {Hopman}, C., \& {Alexander}, T. 2009, \apjl, 698, L64,
  \dodoi{10.1088/0004-637X/698/1/L64}

\bibitem[{{Kormendy}(2004)}]{Kormendy04}
{Kormendy}, J. 2004, in Coevolution of Black Holes and Galaxies, ed. L.~C.
  {Ho}, 1.
\newblock \doarXiv{astro-ph/0306353}

\bibitem[{{Kormendy} \& {Ho}(2013)}]{KormendyHo13}
{Kormendy}, J., \& {Ho}, L.~C. 2013, \araa, 51, 511,
  \dodoi{10.1146/annurev-astro-082708-101811}

\bibitem[{{Kremer} {et~al.}(2022){Kremer}, {Lombardi}, {Lu}, {Piro}, \&
  {Rasio}}]{Kremer+22}
{Kremer}, K., {Lombardi}, James~C., J., {Lu}, W., {Piro}, A.~L., \& {Rasio},
  F.~A. 2022, arXiv e-prints, arXiv:2201.12368.
\newblock \doarXiv{2201.12368}

\bibitem[{{Kremer} {et~al.}(2020){Kremer}, {Spera}, {Becker}, {Chatterjee}, {Di
  Carlo}, {Fragione}, {Rodriguez}, {Ye}, \& {Rasio}}]{Kremer+20}
{Kremer}, K., {Spera}, M., {Becker}, D., {et~al.} 2020, \apj, 903, 45,
  \dodoi{10.3847/1538-4357/abb945}

\bibitem[{{Lai} {et~al.}(1993){Lai}, {Rasio}, \& {Shapiro}}]{Lai+93}
{Lai}, D., {Rasio}, F.~A., \& {Shapiro}, S.~L. 1993, \apj, 412, 593,
  \dodoi{10.1086/172946}

\bibitem[{{Levin} \& {Beloborodov}(2003)}]{Levin+03}
{Levin}, Y., \& {Beloborodov}, A.~M. 2003, \apjl, 590, L33,
  \dodoi{10.1086/376675}

\bibitem[{{Limongi} \& {Chieffi}(2018)}]{LimongiChieffi}
{Limongi}, M., \& {Chieffi}, A. 2018, \apjs, 237, 13,
  \dodoi{10.3847/1538-4365/aacb24}

\bibitem[{{Linial} \& {Sari}(2022)}]{LinialSari22}
{Linial}, I., \& {Sari}, R. 2022, \apj, 940, 101,
  \dodoi{10.3847/1538-4357/ac9bfd}

\bibitem[{{Lombardi} {et~al.}(2002){Lombardi}, {Warren}, {Rasio}, {Sills}, \&
  {Warren}}]{Lombardi+02}
{Lombardi}, James~C., J., {Warren}, J.~S., {Rasio}, F.~A., {Sills}, A., \&
  {Warren}, A.~R. 2002, \apj, 568, 939, \dodoi{10.1086/339060}

\bibitem[{{Lu} \& {Naoz}(2019)}]{Lu+19}
{Lu}, C.~X., \& {Naoz}, S. 2019, \mnras, 484, 1506,
  \dodoi{10.1093/mnras/stz036}

\bibitem[{{Lu} {et~al.}(2013){Lu}, {Do}, {Ghez}, {Morris}, {Yelda}, \&
  {Matthews}}]{Lu+13}
{Lu}, J.~R., {Do}, T., {Ghez}, A.~M., {et~al.} 2013, \apj, 764, 155,
  \dodoi{10.1088/0004-637X/764/2/155}

\bibitem[{{Lu} {et~al.}(2009){Lu}, {Ghez}, {Hornstein}, {Morris}, {Becklin}, \&
  {Matthews}}]{Lu+09}
{Lu}, J.~R., {Ghez}, A.~M., {Hornstein}, S.~D., {et~al.} 2009, \apj, 690, 1463,
  \dodoi{10.1088/0004-637X/690/2/1463}

\bibitem[{{Madigan} {et~al.}(2017){Madigan}, {McCourt}, \&
  {O'Leary}}]{Madigan+17}
{Madigan}, A.-M., {McCourt}, M., \& {O'Leary}, R.~M. 2017, \mnras, 465, 2310,
  \dodoi{10.1093/mnras/stw2815}

\bibitem[{{Mastrobuono-Battisti} {et~al.}(2021){Mastrobuono-Battisti},
  {Church}, \& {Davies}}]{Mastrobuono-Battisti+21}
{Mastrobuono-Battisti}, A., {Church}, R.~P., \& {Davies}, M.~B. 2021, \mnras,
  505, 3314, \dodoi{10.1093/mnras/stab1409}

\bibitem[{{Montero-Casta{\~n}o} {et~al.}(2009){Montero-Casta{\~n}o},
  {Herrnstein}, \& {Ho}}]{Montero-Castano+09}
{Montero-Casta{\~n}o}, M., {Herrnstein}, R.~M., \& {Ho}, P. T.~P. 2009, \apj,
  695, 1477, \dodoi{10.1088/0004-637X/695/2/1477}

\bibitem[{{Morris}(1993)}]{Morris93}
{Morris}, M. 1993, \apj, 408, 496, \dodoi{10.1086/172607}

\bibitem[{{Murphy} {et~al.}(1991){Murphy}, {Cohn}, \& {Durisen}}]{Murphy+91}
{Murphy}, B.~W., {Cohn}, H.~N., \& {Durisen}, R.~H. 1991, \apj, 370, 60,
  \dodoi{10.1086/169793}

\bibitem[{{Murray-Clay} \& {Loeb}(2012)}]{Murray-Clay+12}
{Murray-Clay}, R.~A., \& {Loeb}, A. 2012, Nature Communications, 3, 1049,
  \dodoi{10.1038/ncomms2044}

\bibitem[{{Naoz} {et~al.}(2022){Naoz}, {Rose}, {Michaely}, {Melchor},
  {Ramirez-Ruiz}, {Mockler}, \& {Schnittman}}]{Naoz+22}
{Naoz}, S., {Rose}, S.~C., {Michaely}, E., {et~al.} 2022, \apjl, 927, L18,
  \dodoi{10.3847/2041-8213/ac574b}

\bibitem[{{Nogueras-Lara} {et~al.}(2021){Nogueras-Lara}, {Sch{\"o}del}, \&
  {Neumayer}}]{Nogueras-Lara+21}
{Nogueras-Lara}, F., {Sch{\"o}del}, R., \& {Neumayer}, N. 2021, \apj, 920, 97,
  \dodoi{10.3847/1538-4357/ac185e}

\bibitem[{{O'Leary} {et~al.}(2009){O'Leary}, {Kocsis}, \& {Loeb}}]{OLeary+09}
{O'Leary}, R.~M., {Kocsis}, B., \& {Loeb}, A. 2009, \mnras, 395, 2127,
  \dodoi{10.1111/j.1365-2966.2009.14653.x}

\bibitem[{{Owen} \& {Lin}(2023)}]{Owen+23}
{Owen}, J.~E., \& {Lin}, D. N.~C. 2023, \mnras, 519, 397,
  \dodoi{10.1093/mnras/stac3506}

\bibitem[{{Paumard} {et~al.}(2006){Paumard}, {Genzel}, {Martins}, {Nayakshin},
  {Beloborodov}, {Levin}, {Trippe}, {Eisenhauer}, {Ott}, {Gillessen}, {Abuter},
  {Cuadra}, {Alexander}, \& {Sternberg}}]{Paumard+06}
{Paumard}, T., {Genzel}, R., {Martins}, F., {et~al.} 2006, \apj, 643, 1011,
  \dodoi{10.1086/503273}

\bibitem[{{Pfuhl} {et~al.}(2011){Pfuhl}, {Fritz}, {Zilka}, {Maness},
  {Eisenhauer}, {Genzel}, {Gillessen}, {Ott}, {Dodds-Eden}, \&
  {Sternberg}}]{Pfuhl+11}
{Pfuhl}, O., {Fritz}, T.~K., {Zilka}, M., {et~al.} 2011, \apj, 741, 108,
  \dodoi{10.1088/0004-637X/741/2/108}

\bibitem[{{Prodan} {et~al.}(2015){Prodan}, {Antonini}, \& {Perets}}]{Prodan+15}
{Prodan}, S., {Antonini}, F., \& {Perets}, H.~B. 2015, \apj, 799, 118,
  \dodoi{10.1088/0004-637X/799/2/118}

\bibitem[{{Rauch}(1999)}]{Rauch99}
{Rauch}, K.~P. 1999, \apj, 514, 725, \dodoi{10.1086/306953}

\bibitem[{{Renzo} {et~al.}(2020){Renzo}, {Farmer}, {Justham}, {G{\"o}tberg},
  {de Mink}, {Zapartas}, {Marchant}, \& {Smith}}]{Renzo+20}
{Renzo}, M., {Farmer}, R., {Justham}, S., {et~al.} 2020, \aap, 640, A56,
  \dodoi{10.1051/0004-6361/202037710}

\bibitem[{{Rodriguez} {et~al.}(2022){Rodriguez}, {Weatherford}, {Coughlin},
  {Amaro-Seoane}, {Breivik}, {Chatterjee}, {Fragione}, {K{\i}ro{\u{g}}lu},
  {Kremer}, {Rui}, {Ye}, {Zevin}, \& {Rasio}}]{Rodriguez+22}
{Rodriguez}, C.~L., {Weatherford}, N.~C., {Coughlin}, S.~C., {et~al.} 2022,
  \apjs, 258, 22, \dodoi{10.3847/1538-4365/ac2edf}

\bibitem[{{Rose} {et~al.}(2020){Rose}, {Naoz}, {Gautam}, {Ghez}, {Do}, {Chu},
  \& {Becklin}}]{Rose+20}
{Rose}, S.~C., {Naoz}, S., {Gautam}, A.~K., {et~al.} 2020, \apj, 904, 113,
  \dodoi{10.3847/1538-4357/abc557}

\bibitem[{{Rose} {et~al.}(2022){Rose}, {Naoz}, {Sari}, \& {Linial}}]{Rose+22}
{Rose}, S.~C., {Naoz}, S., {Sari}, R., \& {Linial}, I. 2022, \apjl, 929, L22,
  \dodoi{10.3847/2041-8213/ac6426}

\bibitem[{{Rubin} \& {Loeb}(2011)}]{RubinLoeb}
{Rubin}, D., \& {Loeb}, A. 2011, Advances in Astronomy, 2011, 174105,
  \dodoi{10.1155/2011/174105}

\bibitem[{{Schartmann} {et~al.}(2012){Schartmann}, {Burkert}, {Alig},
  {Gillessen}, {Genzel}, {Eisenhauer}, \& {Fritz}}]{Schartmann+12}
{Schartmann}, M., {Burkert}, A., {Alig}, C., {et~al.} 2012, \apj, 755, 155,
  \dodoi{10.1088/0004-637X/755/2/155}

\bibitem[{{Sch{\"o}del} {et~al.}(2014){Sch{\"o}del}, {Feldmeier}, {Kunneriath},
  {Stolovy}, {Neumayer}, {Amaro-Seoane}, \& {Nishiyama}}]{Schodel+14}
{Sch{\"o}del}, R., {Feldmeier}, A., {Kunneriath}, D., {et~al.} 2014, \aap, 566,
  A47, \dodoi{10.1051/0004-6361/201423481}

\bibitem[{{Sch{\"o}del} {et~al.}(2018){Sch{\"o}del}, {Gallego-Cano}, {Dong},
  {Nogueras-Lara}, {Gallego-Calvente}, {Amaro-Seoane}, \&
  {Baumgardt}}]{Schodel+18}
{Sch{\"o}del}, R., {Gallego-Cano}, E., {Dong}, H., {et~al.} 2018, \aap, 609,
  A27, \dodoi{10.1051/0004-6361/201730452}

\bibitem[{{Sch{\"o}del} {et~al.}(2003){Sch{\"o}del}, {Genzel}, {Ott}, \&
  {Eckart}}]{schodel+03}
{Sch{\"o}del}, R., {Genzel}, R., {Ott}, T., \& {Eckart}, A. 2003, Astronomische
  Nachrichten Supplement, 324, 535, \dodoi{10.1002/asna.200385048}

\bibitem[{{Sch{\"o}del} {et~al.}(2020){Sch{\"o}del}, {Nogueras-Lara},
  {Gallego-Cano}, {Shahzamanian}, {Gallego-Calvente}, \&
  {Gardini}}]{Schodel+20}
{Sch{\"o}del}, R., {Nogueras-Lara}, F., {Gallego-Cano}, E., {et~al.} 2020,
  arXiv e-prints, arXiv:2007.15950.
\newblock \doarXiv{2007.15950}

\bibitem[{{Sills} {et~al.}(2001){Sills}, {Faber}, {Lombardi}, {Rasio}, \&
  {Warren}}]{Sills+01}
{Sills}, A., {Faber}, J.~A., {Lombardi}, James~C., J., {Rasio}, F.~A., \&
  {Warren}, A.~R. 2001, \apj, 548, 323, \dodoi{10.1086/318689}

\bibitem[{{Sills} {et~al.}(1997){Sills}, {Lombardi}, {Bailyn}, {Demarque},
  {Rasio}, \& {Shapiro}}]{Sills+97}
{Sills}, A., {Lombardi}, James~C., J., {Bailyn}, C.~D., {et~al.} 1997, \apj,
  487, 290, \dodoi{10.1086/304588}

\bibitem[{{Spera} \& {Mapelli}(2017)}]{SperaMapelli17}
{Spera}, M., \& {Mapelli}, M. 2017, \mnras, 470, 4739,
  \dodoi{10.1093/mnras/stx1576}

\bibitem[{{Stephan} {et~al.}(2016){Stephan}, {Naoz}, {Ghez}, {Witzel},
  {Sitarski}, {Do}, \& {Kocsis}}]{Stephan+16}
{Stephan}, A.~P., {Naoz}, S., {Ghez}, A.~M., {et~al.} 2016, ArXiv e-prints.
\newblock \doarXiv{1603.02709}

\bibitem[{{Stephan} {et~al.}(2019{\natexlab{a}}){Stephan}, {Naoz}, {Ghez},
  {Morris}, {Ciurlo}, {Do}, {Breivik}, {Coughlin}, \& {Rodriguez}}]{Stephan+19}
---. 2019{\natexlab{a}}, \apj, 878, 58, \dodoi{10.3847/1538-4357/ab1e4d}

\bibitem[{{Stephan} {et~al.}(2019{\natexlab{b}}){Stephan}, {Naoz}, {Ghez},
  {Morris}, {Ciurlo}, {Do}, {Breivik}, {Coughlin}, \& {Rodriguez}}]{Stephan19}
---. 2019{\natexlab{b}}, arXiv e-prints, arXiv:1903.00010.
\newblock \doarXiv{1903.00010}

\bibitem[{{Torricelli-Ciamponi} {et~al.}(2000){Torricelli-Ciamponi}, {Foellmi},
  {Courvoisier}, \& {Paltani}}]{Torricelli-C+00}
{Torricelli-Ciamponi}, G., {Foellmi}, C., {Courvoisier}, T.~J.~L., \&
  {Paltani}, S. 2000, \aap, 358, 57

\bibitem[{{Tremaine} {et~al.}(2002){Tremaine}, {Gebhardt}, {Bender}, {Bower},
  {Dressler}, {Faber}, {Filippenko}, {Green}, {Grillmair}, {Ho}, {Kormendy},
  {Lauer}, {Magorrian}, {Pinkney}, \& {Richstone}}]{Tremaine+02}
{Tremaine}, S., {Gebhardt}, K., {Bender}, R., {et~al.} 2002, \apj, 574, 740,
  \dodoi{10.1086/341002}

\bibitem[{{Witzel} {et~al.}(2014){Witzel}, {Ghez}, {Morris}, {Sitarski},
  {Boehle}, {Naoz}, {Campbell}, {Becklin}, {Canalizo}, {Chappell}, {Do}, {Lu},
  {Matthews}, {Meyer}, {Stockton}, {Wizinowich}, \& {Yelda}}]{Witzel+14}
{Witzel}, G., {Ghez}, A.~M., {Morris}, M.~R., {et~al.} 2014, \apjl, 796, L8,
  \dodoi{10.1088/2041-8205/796/1/L8}

\bibitem[{{Witzel} {et~al.}(2017){Witzel}, {Sitarski}, {Ghez}, {Morris},
  {Hees}, {Do}, {Lu}, {Naoz}, {Boehle}, {Martinez}, {Chappell}, {Sch{\"o}del},
  {Meyer}, {Yelda}, {Becklin}, \& {Matthews}}]{Witzel+17}
{Witzel}, G., {Sitarski}, B.~N., {Ghez}, A.~M., {et~al.} 2017, \apj, 847, 80,
  \dodoi{10.3847/1538-4357/aa80ea}

\bibitem[{{Woosley}(2017)}]{Woosley+17}
{Woosley}, S.~E. 2017, \apj, 836, 244, \dodoi{10.3847/1538-4357/836/2/244}

\bibitem[{{Zaja{\v{c}}ek} {et~al.}(2020){Zaja{\v{c}}ek}, {Araudo}, {Karas},
  {Czerny}, \& {Eckart}}]{Zajacek+20}
{Zaja{\v{c}}ek}, M., {Araudo}, A., {Karas}, V., {Czerny}, B., \& {Eckart}, A.
  2020, \apj, 903, 140, \dodoi{10.3847/1538-4357/abbd94}

\bibitem[{{Zaja{\v{c}}ek} {et~al.}(2017){Zaja{\v{c}}ek}, {Britzen}, {Eckart},
  {Shahzamanian}, {Busch}, {Karas}, {Parsa}, {Peissker}, {Dov{\v{c}}iak},
  {Subroweit}, {Dinnbier}, \& {Zensus}}]{Zajacek+17}
{Zaja{\v{c}}ek}, M., {Britzen}, S., {Eckart}, A., {et~al.} 2017, \aap, 602,
  A121, \dodoi{10.1051/0004-6361/201730532}

\end{thebibliography}
\bibliographystyle{aasjournal}



\end{document}